\documentclass[apj,numberedappendix,revtex4]{emulateapj}
\usepackage[backref,breaklinks,colorlinks,citecolor=blue]{hyperref}
\usepackage{amssymb,amsmath}
\usepackage{graphicx}
\usepackage{appendix}
\usepackage{natbib}
\usepackage{enumerate}
\usepackage{multirow}
\usepackage{bm}
\usepackage{mathrsfs}

\newcommand{\etal}{et~al.~}

\altaffiltext{\MIT}{Department of Physics, Massachusetts Institute of Technology, Cambridge, MA 02139, USA}
\altaffiltext{\MKI}{Kavli Institute for Astrophysics and Space Research, Massachusetts Institute of Technology, 77 Massachusetts Avenue, Cambridge, MA 02139}
\altaffiltext{\Waterlooa}{Department of Physics and Astronomy, University of Waterloo, 200 University Avenue West, Waterloo, ON, N2L 3G1, Canada}
\altaffiltext{\Waterloob}{Waterloo Centre for Astrophysics, University of Waterloo, 200 University Avenue West, Waterloo, ON, N2L 3G1, Canada}
\altaffiltext{\Waterlooc}{Perimeter Institute for Theoretical Physics, 31 Caroline St N, Waterloo, ON, N2L 2Y5, Canada}
\altaffiltext{\MSFC}{Marshall Space Flight Center, Huntsville, AL 35811, USA}
\altaffiltext{\INAFB}{INAF, Osservatorio di Astrofisica e Scienza dello Spazio, via Piero Gobetti 93/3, 40129 Bologna, Italy}
\altaffiltext{\Princeton}{Department of Astrophysical Sciences, Princeton University, 4 Ivy Lane, Princeton, NJ 08544-1001, USA}
\altaffiltext{\Dartmouth}{Dartmouth College, 6127 Wilder Laboratory, Hanover, NH, 03755 USA}
\altaffiltext{\Queens}{Department of Physics, Engineering Physics and Astronomy, Queen's University, Kingston, ON K7L 3N6, Canada}

\def\MIT{1}
\def\MKI{2}
\def\Waterlooa{3}
\def\Waterloob{4}
\def\Waterlooc{5}
\def\MSFC{6}
\def\INAFB{7}
\def\Princeton{8}
\def\Dartmouth{9}
\def\Queens{10}

\begin{document}


\title{
Observational Evidence for Enhanced Black Hole Accretion in Giant Elliptical Galaxies
}
   
\author{Michael McDonald\altaffilmark{\MIT,\MKI},
Brian R.~McNamara\altaffilmark{\Waterlooa,\Waterloob,\Waterlooc},
Michael S.~Calzadilla \altaffilmark{\MIT,\MKI},
Chien-Ting Chen\altaffilmark{\MSFC},
\\
Massimo~Gaspari\altaffilmark{\INAFB,\Princeton},
Ryan C.~Hickox\altaffilmark{\Dartmouth},
Erin Kara\altaffilmark{\MIT,\MKI},
Ilia Korchagin\altaffilmark{\Queens}
}

\email{Email: mcdonald@space.mit.edu}   

 
\begin{abstract}

We present a study of the relationship between black hole accretion rate (BHAR) and star formation rate (SFR) in a sample of giant elliptical galaxies. These galaxies, which live at the centers of galaxy groups and clusters, have star formation and black hole activity that is primarily fueled by gas condensing out of the hot intracluster medium. For a sample of 46 galaxies spanning 5 orders of magnitude in BHAR and SFR, we find a mean ratio of $\log_{10}(\textrm{BHAR/SFR})=-1.45\pm0.2$, independent of the methodology used to constrain both SFR and BHAR. This ratio is significantly higher than most previously-published values for field galaxies. We investigate whether these high BHAR/SFR ratios are driven by high BHAR, low SFR, or a different accretion efficiency in radio galaxies. The data suggest that the high BHAR/SFR ratios are primarily driven by boosted black hole accretion in spheroidal galaxies compared to their disk counterparts. We propose that angular momentum of the cool gas is the primary driver in suppressing BHAR in lower mass galaxies, with massive galaxies accreting gas that has condensed out of the hot phase on nearly radial trajectories. 
Additionally, we demonstrate that the relationship between \emph{specific} BHAR and SFR has much less scatter over 6 orders of magnitude in both parameters, due to competing dependence on morphology between the M$_{\textrm{BH}}$--M$_*$ and BHAR--SFR relations. In general, active galaxies selected by typical techniques have sBHAR/sSFR $\sim$ 10, while galactic nuclei with no clear AGN signatures have sBHAR/sSFR $\sim$ 1, consistent with a universal M$_{\textrm{BH}}$--M$_{\textrm{spheroid}}$ relation.


\end{abstract}

\keywords{accretion -- galaxies: active -- galaxies: elliptical and lenticular, cD -- galaxies: jets -- galaxies: Seyfert --  galaxies: quasars: general}

\section{Introduction}
\setcounter{footnote}{0}


Giant elliptical galaxies, which typically occupy the centers of galaxy groups and clusters, are the most massive galaxies in the Universe. While already exceptionally massive \citep[M$_*$ $\sim$10$^{12}$ M$_{\odot}$;][]{lidman12}, these galaxies are orders of magnitude less massive (and luminous) than early galaxy formation simulations predicted \citep[see review by ][]{silk12}. 
It is currently thought that energetic feedback from accreting supermassive black holes, or active galactic nuclei (AGN), is responsible for preventing cooling of intergalactic gas on large physical scales \citep[see reviews by][]{mcnamara07,mcnamara12,fabian12,gaspari20} and over most of cosmic time \citep[see e.g.,][]{mcdonald13b,hlavacek15,mcdonald17} in the most massive halos (galaxy groups and clusters). In massive galaxies, this feedback is most commonly ``radio-mode'', which consists of relativistic jets injecting energy into the surrounding medium primarily via mechanical means (inflating bubbles/cavities). This inflation of bubbles via radio jets can inject $\sim$10$^{42}$--10$^{46}$ ergs s$^{-1}$ of energy into the surrounding hot halo, on par with the cooling luminosity of the intragroup or intracluster medium \citep[e.g.,][]{rafferty06,hlavacek12,hlavacek15}. While the mechanism for coupling this energy to the hot phase is currently poorly understood, these powerful jets appear able to suppress cooling by two orders of magnitude, with typical star formation rates in central brightest cluster galaxies (BCGs) being only 1\% of the predicted cooling rate based on the amount and temperature of intracluster gas \citep[e.g.,][]{odea08,mcdonald18a}.

AGN feedback is certainly not limited to only the most massive galaxies, though that is where the effects are most dramatic. Given that supermassive black holes (SMBHs) are ubiquitous at the centers of galaxies \citep[e.g.,][]{kormendy95,magorrian98,kormendy13}, one may expect to see AGN in nearly every galaxy as well. %
%
%
The fact that we observe luminous AGN in $\ll$100\% of galaxies implies that accretion is not continuous, and that SMBHs go through active and inactive periods, with X-ray duty cycles of $\sim$1\% in typical galaxies at $z<1$ \citep[e.g.,][]{,delvecchio20}. However, this duty cycle is strongly dependent on the host galaxy properties and the threshold for the term ``active'', with $\sim$100\% of the most massive (M$_*$ $>$ 10$^{11}$ M$_{\odot}$) galaxies harboring (at minimum) low-luminosity radio sources, corresponding to SMBHs accreting at $>$10$^{-7}$$\dot{M}_{Edd}$ \citep{sabater19}. %
%
%
In general, black hole accretion rates (BHARs) and duty cycles appear to correlate with the host galaxy star formation rate (SFR), with the most star-forming galaxies typically harboring the most rapidly accreting black holes \citep[e.g.,][]{diamond-stanic12,chen13, drouart14,delvecchio15,gurkan15,rodighiero15,xu15,dong16,dai18,yang19b,zhuang20}. There is significant scatter in these relations, largely due to the fact that AGN luminosities can vary on a wide range of timescales from hours to Myr, significantly shorter than the typical $\sim$100\,Myr timescale of star formation \citep{hickox14}.
The mean ratio of the BHAR to the SFR is highly variable across these studies, with estimates ranging from BHAR/SFR $\sim$ 1/300 \citep{diamond-stanic12,xu15,dong16,yang19b} to BHAR/SFR $\sim$ 1/5000 \citep{chen13,delvecchio15,rodighiero15}.
On the surface, the correlation between SFR and BHAR is unsurprising: both star formation and black hole accretion are fueled by gas, so more gas-rich galaxies ought to have elevated SFR and BHAR, while gas-poor galaxies ought to have suppressed SFR and BHAR. Such a correlation is probably also necessary to arrive at the observed trends between host galaxy properties and SMBH mass \citep[e.g.,][]{kormendy13,mcconnell13}, in particular those between the stellar mass of the host galaxy and the black hole mass \citep[e.g.,][]{reines15}. However, there is evidence that the relationship between SFR and BHAR is not universal, with some authors finding weaker or non-existent correlations when selecting on different galaxy/AGN types \citep[e.g.,][]{stanley15,shimizu17,yang17,yang19b}.


Despite significant efforts by the community in studying the properties of radio-mode AGN at the centers of groups and clusters \citep[e.g.,][]{dunn06,rafferty06,best07,mcnamara07,rafferty08,cavagnolo10,birzan12,mcnamara12,russell13,hlavacek15}, the relationship between SFR and BHAR has not been studied in these systems. Giant elliptical galaxies in massive halos provide a unique opportunity to investigate this correlation, as their time-averaged powers can be inferred from their influence on the surrounding hot halo. While the majority of previous studies have used X-ray luminosity, optical emission lines, or mid-IR luminosity as a proxy for the power output of an AGN, we can instead measure the mechanical power output of radio-loud AGN by calculating the work required to inflate the observed bubbles in the intracluster or intragroup medium \citep[e.g.,][]{birzan04,dunn05,rafferty06,hlavacek12}.  Such measurements of the jet power are sensitive to accretion rates as low as $<$10$^{-5}$ Eddington \citep{russell13}, a regime that proxies based on the luminosity of the accretion disk are generally insensitive to. Further, estimates of the jet power based on X-ray cavities provide time-averaged estimates of the accretion rate on tens of Myr timescales, similar to standard SFR indicators, avoiding the complication of comparing instantaneous and time-averaged quantities. Finally, giant elliptical galaxies are (for the most part) quite passive, allowing us to study the BHAR--SFR relation for the first time in galaxies with specific star formation rates ($sSFR \equiv SFR/M_*$) as low as 10$^{-4}$ Gyr$^{-1}$.

In this work, we provide the first assessment of the BHAR--SFR relation in giant elliptical galaxies. We focus on the sample of \cite{russell13}, which contains a wide variety of galaxies, groups, and clusters, with accretion rates ranging from $<$10$^{-5}$ $\dot{M}_{Edd}$ to $\sim$ $\dot{M}_{Edd}$. The properties of this sample, and our methodology for measuring BHAR and SFR are described in \S2.  We present the BHAR--SFR relation for giant elliptical galaxies in \S3, and consider the dependence of this relation on the stellar mass of the host galaxy. In \S4 we compare our findings to the literature, and attempt to determine the primary physical drivers for the observed BHAR/SFR ratios. Finally, in \S5 we provide a unified picture of the BHAR--SFR relation across all galaxy and AGN types, and discuss the connection between this relation and the relationship between SMBH and host galaxy mass.

Throughout this work we assume $\Lambda$CDM cosmology with H$_0$ = 70 km s$^{-1}$ Mpc$^{-1}$, $\Omega_M$ = 0.3, $\Omega_{\Lambda}$ = 0.7. Unless otherwise stated, quoted scatters and uncertainties are 1$\sigma$ RMS. 

\section{Data}

\begin{deluxetable*}{cccccccc}[p]
\tabletypesize{\footnotesize}
\tablecolumns{7}
\tablewidth{0pt}
\tablecaption{AGN Power and Cool/Cold Gas Supply in Giant Elliptical Galaxies \label{table:data}}
\tablehead{
\colhead{Name} & \colhead{$z$} & \colhead{P$_{cav}$} & \colhead{L$_{H\alpha}$} & \colhead{L$_{H\alpha,corr}$} & \colhead{Ref} & \colhead{M$_{H_2}$} & \colhead{Ref}\\ 
 & & \colhead{[$10^{44}$ erg s$^{-1}$]} & \colhead{[$10^{40}$ erg s$^{-1}$]} &  \colhead{[$10^{40}$ erg s$^{-1}$]} & & \colhead{[$10^8$ M$_{\odot}$]} & }
\startdata
2A0335+096 & 0.0349 & 0.24$_{-0.09}^{+0.24}$ & $<$2.58 & $<$2.58 &  2 & 17.0 $\pm$ 5.40 & 14 \\
3C295 	    & 0.4641 & 0.36$_{-0.15}^{+0.15}$ & $<$232 & $<$232 & 2 & -- & -- \\
3C388 	    & 0.0917 & 2.00$_{-0.97}^{+2.85}$ & $<$6.76 & $<$6.76 & 1 & $<$12 & 17 \\
4C55.16       & 0.2411 & 4.19$_{-2.00}^{+4.55}$ & $<$800 &  $<$800 & 2 & $<$160 & 14 \\
ABELL 0085 & 0.0551 & 0.37$_{-0.16}^{+0.39}$ & 1.60 & 0.95 & 5 & 4.50 $\pm$ 2.50 & 14 \\
ABELL 0133 & 0.0566 & 6.21$_{-1.80}^{+3.16}$ & 1.20 & 0.76 & 5 & -- & -- \\
ABELL 0262 & 0.0166 & 0.10$_{-0.03}^{+0.08}$ & 2.58 & 2.35 & 6 & 4.00 $\pm$ 1.30 & 14 \\
ABELL 0478 & 0.0881 & 1.00$_{-0.37}^{+0.86}$ & 23.0 & 22.5 & 5 & 19.0 $\pm$ 12.0 & 14 \\
ABELL 1795 & 0.0625 & 1.60$_{-0.70}^{+2.36}$ & 26.7 & 26.2 & 7 & 48.00 $\pm$ 6.00 & 14 \\
ABELL 1835 & 0.2523 & 14.3$_{5.07}^{+14.3}$ & 785 & 784 & 3 & 501 $\pm$ 231 & 11 \\
ABELL 2029 & 0.0773 & 0.87$_{-0.32}^{+0.59}$ & $<$0.44 & $<$0.44 & 7 & $<$17 & 14 \\
ABELL 2052 & 0.0351 & 1.50$_{-0.43}^{+2.05}$ & 1.80 & 1.37 & 5 & 9.00 $\pm$ 3.60 & 14 \\
ABELL 2199 & 0.0302 & 2.70$_{-0.98}^{+2.62}$ & 1.42 & 1.31 & 2 & $<$2.6 & 14 \\
ABELL 2390 & 0.2280 & 100$_{-97.2}^{+107}$ & 109 & 108 & 5 & $<$180 & 14 \\
ABELL 2597 & 0.0852 & 0.67$_{-0.34}^{+0.89}$ & 53.8 & 53.7 & 2 & 26.0 $\pm$ 13.0 & 14 \\
ABELL 4059 & 0.0475 & 0.96$_{-0.46}^{+0.94}$ & 4.10 & 3.51 & 5 & -- & -- \\
CENTAURUS  & 0.0114 & 0.07$_{-0.03}^{+0.06}$ & 3.53 & 3.17&  6 & $<$2.2 & 14 \\
CYGNUS A     & 0.0561 & 13.0$_{-3.92}^{+11.5}$ & $<$21.3 & $<$21.3 & 2 & 10.0 $\pm$ 3.80 & 14 \\
HCG 0062      & 0.0137 & 0.04$_{-0.02}^{+0.06}$ & 0.12 & $<$0.12 & 6 & -- & -- \\
HERCULES A & 0.1550 & 3.11$_{-1.31}^{+4.12}$ & 1.30 & 1.18 & 2 & -- & -- \\
HYDRA A        & 0.0549 & 4.29$_{-1.22}^{+2.28}$ & 13.0 & 12.7 & 5 & 39.0 $\pm$ 16.0 & 14 \\
M84   	      & 0.0035 & 0.01$_{-0.01}^{+0.02}$ & 0.33 & 0.16 & 6 & 0.03 $\pm$ 0.01 & 13 \\
M89 	              & 0.0011 & 0.02$_{-0.01}^{+0.01}$ & 0.18 & 0.11 & 6 & $<$0.19 & 10 \\
MKW3S 	     & 0.0442 & 4.10$_{-1.07}^{+4.31}$ & 0.89 & 0.84 & 1 & $<$5.2 & 15 \\
MS 0735.6+7421 & 0.2160 & 60.7$_{-20.3}^{+20.4}$ & 124 & 124 & 2 & $<$30 & 16 \\
NGC 1316       & 0.0059 & 0.01$_{-0.01}^{+0.00}$ & 0.27 & $<$0.27 & 6 & 5.00 $\pm$ 3.45 & 8 \\
NGC 1600       & 0.0156 & 0.02$_{-0.01}^{+0.01}$ & 0.30 & $<$0.30 & 6 & $<$2.58 & 19 \\
NGC 4261       & 0.0075 & 0.10$_{-0.05}^{+0.05}$ & 0.04 & $<$0.04 & 6 & $<$0.48 & 10 \\
NGC 4472       & 0.0033 & 0.01$_{-0.00}^{+0.00}$ & 0.40 & 0.17 & 6 & 0.40 $\pm$ 0.18 & 9 \\
NGC 4636       & 0.0031 & 0.03$_{-0.01}^{+0.01}$ & 0.49 & 0.47 & 6 & $<$0.07 & 10 \\
NGC 4782       & 0.0154 & 0.02$_{-0.02}^{+0.01}$ & 0.78 & 0.36 & 6 & -- & -- \\
NGC 5044       & 0.0093 & 0.04$_{-0.02}^{+0.01}$ & 0.54 & 0.44 & 5 & 2.30 $\pm$ 0.83 & 14 \\
NGC 5813       & 0.0066 & 0.02$_{-0.00}^{+0.00}$ & 0.04 & $<$0.04 & 5 & $<$0.49 & 10 \\
NGC 5846       & 0.0057 & 0.01$_{-0.01}^{+0.00}$ & 0.08 & 0.04 & 6 & $<$0.6 & 10 \\
NGC 6269       & 0.0348 & 0.02$_{-0.01}^{+0.01}$ & -- & -- & -- & -- & -- \\
NGC 6338       & 0.0274 & 0.11$_{-0.07}^{+0.04}$ & 2.94 & 2.64 & 4 & -- & -- \\
PKS 0745-191   & 0.1028 & 17.0$_{-6.33}^{+15.1}$ & 28.0 & 27.8 & 2 & 40.0 $\pm$ 9.00 & 14 \\
PKS 1404-267   & 0.0218 & 0.20$_{-0.10}^{+0.26}$ & -- & -- & -- & 3.30 $\pm$ 1.50 & 14 \\
RXC J0352.9+1941 & 0.109  & 0.96$_{-0.30}^{+0.35}$ & 62.0 & 61.8 & 5 & 49.0 $\pm$ 19.0 & 14 \\
RXC J1459.4-1811 & 0.2357 & 11.8$_{-4.66}^{+5.16}$ & 241 & 240 & 5 & 220 $\pm$ 110 & 14 \\
RXC J1524.2-3154 & 0.1028 & 1.07$_{-0.35}^{+0.38}$ & 45.9 & 45.6 & 5 & 29.0 $\pm$ 16.0 & 14 \\
RXC J1558.3-1410 & 0.0970 & 4.60$_{-1.79}^{+2.16}$ & 22.0 & 21.3& 5 & 53.0 $\pm$ 19.0 & 14 \\
Sersic 159-03 	 & 0.0580 & 7.79$_{-3.46}^{+8.50}$ & 11.5 & 11.1 & 7 & -- & -- \\
UGC 00408 	 & 0.0147 & 0.04$_{-0.03}^{+0.03}$ & -- & -- & -- & $<$1.35 & 12 \\
Zwicky 2701 	 & 0.2150 & 5.71$_{-2.11}^{+2.28}$ & 5.41 & 5.30 & 2 & -- & -- \\
Zwicky 3146 	 & 0.2906 & 58.1$_{-26.5}^{+71.5}$ & 582 & 583 & 2 & 560 $\pm$ 200 & 14
\enddata
\tablecomments{Primary sample of 46 galaxies/groups/clusters, drawn from \cite{russell13}. When names of groups/clusters are given, we are referring to the central, radio-loud galaxy. Redshifts are from the NASA/IPAC Extragalactic Database and cavity powers are from \cite{russell13}. Corrected H$\alpha$ luminosities are quoted, for which we have removed the contribution from evolved stellar populations (\S2.3). References for H$\alpha$ luminosities: 1:~\cite{buttiglione09}; 2:~\cite{cavagnolo09}; 3:~\cite{crawford99}; 4:~\cite{gomes16}; 5:~\cite{hamer16}; 6:~\cite{lakhchaura18}; 7:~\cite{mcdonald10}. 
References for molecular gas masses: 8:~\cite{horellou01}; 9:~\cite{huchtmeier94}; 10:~\cite{kokusho19}; 11:~\cite{mcnamara14}; 12:~\cite{osullivan15}; 13:~\cite{ocana10}; 14:~\cite{pulido18}; 15:~\cite{salome03}; 16:~\cite{salome08}; 17:~\cite{smolcic11}; 18:~\cite{sofue93}; 
}
\end{deluxetable*}

\subsection{Samples}

Our primary sample is drawn from \cite{russell13}, which consists of 46 galaxies, groups, and clusters of galaxies, spanning a large range in halo mass and AGN power. This sample was selected on the presence of X-ray cavities. Each of these systems has sufficiently deep X-ray data from \emph{Chandra} to infer the total jet power, P$_{cav}$ -- these values and their uncertainties are reported in \cite{russell13} and quoted in Table \ref{table:data}. To this sample we add three additional clusters with rapidly-accreting central galaxies: H1821+643 \citep{russell10}, IRAS09104+4109 \citep{osullivan12}, and Phoenix \citep{mcdonald12c,mcdonald19}. The AGN output in these three systems is predominantly radiative, rather than mechanical.  The addition of these three systems increases the dynamic range of our sample to 6 orders of magnitude in both AGN power and star formation rate.

The sample of \cite{russell13} was selected to span a broad range in black hole accretion rate and, thus, is neither complete nor unbiased. It does, however, probe a broad range of environments, BCG stellar populations, and AGN powers. It is worth noting that, while this sample is biased towards AGN activity, there was no consideration of the BCG SFR in the selection process. While the bulk of our analysis will focus on this sample, where the data quality and dynamic range is superb, we will supplement our analysis with the complete sample of local BCGs from \cite{lauer14} to aid in the discussion (\S6). From this sample of 433 BCGs at $z<0.08$, we draw a representative subsample of 68 BCGs based on the overlapping footprints of the NRAO FIRST survey \citep{becker95} and the Sloan Digital Sky Survey Data Release 8 \citep[SDSS DR8;][]{aihara11}. The addition of this sample, which should be significantly less biased towards active black holes, will allow us to assess the effects of selection bias and non-detections. 

\subsection{Inferring BHAR via P$_{cav}$, L$_{nuc}$, $\dot{\textrm{M}}_{Bondi}$}
The majority of the BHARs used in this work come directly from the cavity powers quoted in \cite{russell13}. We assume $P_{cav} = \epsilon_{acc}\dot{M}_{BH}c^2$, where $\epsilon_{acc}$ is the efficiency with which accreting matter is turned into energy and $\dot{M}_{BH}$ is the accretion rate averaged on large ($>$10 Myr) timescales (i.e., bubble rise times). This assumes that the mechanical power of the AGN dominates the radiative power, which is the case for the 46 galaxies that comprise our main sample \citep[see Figure 12 of][]{russell13}. We assume $\epsilon_{acc} = 0.1$ throughout this work in order to be consistent with the literature, but we will discuss the effects of varying $\epsilon_{acc}$ in \S5.3. In particular, we recognize that $\epsilon_{acc} > 0.1$ is predicted for magnetically-arrested discs, with $\epsilon_{acc}>1$ being possible for maximally-spinning black holes where the jets are tapping into the spin of the black hole to boost the energy output \citep{narayan03, tchekhovskoy11}. We will expand on this in \S5.3.

We also consider BHARs derived based on the bolometric luminosity of the central point source for three radiatively-efficient AGN at the centers of massive clusters. For H1821+643 and the Phoenix cluster, we measure the rest-frame, unabsorbed 2--10\,keV luminosity, and convert to a bolometric luminosity following \cite{hopkins07}. We apply a factor of 0.7 correction to this bolometric luminosity, to correct for the ``double-counting'' of IR reprocessed emission, following \cite{merloni13} and \cite{yang19b}. This yields bolometric corrections of 108 and 99 for H1821+643 and Phoenix, respectively. For IRAS09104+4109, we take the bolometric luminosity directly from \cite{vignali11}, who combine X-ray and infrared data to model the full spectral energy distribution. The implied L$_{2-10 keV}$ bolometric correction for this system is $\sim$150. In all three of these systems, the radiative power output from the central AGN is significantly higher than the mechanical power output, so we assume $\dot{M}_{BH} = L_{bol,nuc}/\epsilon c^2$,

Finally, we consider the Bondi accretion rate \citep[$\dot{M}_{Bondi}=\pi \rho G^2 M^2 c_s^{-3}$;][]{bondi52} for a subsample of 12 systems, taken from \citep{russell13}. These are the only systems for which the data quality is sufficient to constrain the thermodynamics of the hot gas within several Bondi radii and, thus, provide a meaningful estimate of the Bondi rate. We will not incorporate these estimates into our analysis, but simply use them to illustrate an upper limit on the hot mode accretion rate in these systems, when angular momentum, cooling/condensation, and feedback are all neglected.


\subsection{Inferring SFR via L$_{H\alpha}$ and M$_{H_2}$}

There is a wide variety of ways to infer the star formation rate for a galaxy based on broadband, narrowband, and spectroscopic measurements \citep[see review by][]{kennicutt98}. In this work we focus on two gas-based indicators of star formation: the H$\alpha$ emission line and the total molecular gas mass (as probed by the CO (1-0) rotational transition line). Alternative methods include the UV continuum, the mid-IR continuum, and the far-IR continuum, which probe young stars, warm dust, and cold dust, respectively.  We are particularly interested in probing to very low \emph{specific} star formation rates ($\textrm{sSFR} \equiv \textrm{SFR}/\textrm{M}_*$), where we expect to find passive, giant elliptical galaxies. At these very low values of sSFR, all of the traditional indicators of star formation can be contaminated by evolved stars. For example, planetary nebulae and supernova remnants exhibit strong emission lines and can be quite dusty \citep[e.g.,][]{fesen85,bhattacharya19}, while hot, low-mass evolved stars can lead to significant amounts of UV emission and ionization when integrated over a massive stellar population \citep{cidfernandes11,yi11}. While these may seem inconsequential, they can likely dominate the flux in any classical indicator of star formation when sSFR goes below $\sim$10$^{-4}$ Gyr$^{-1}$, or SFR $<$ 0.01 M$_{\odot}$ yr$^{-1}$ for M$_* = 10^{12}$ M$_{\odot}$. For this reason, we have chosen two indicators for which we can either correct for this contamination (L$_{H\alpha}$) or where it should be negligible (M$_{H_2}$).

For 43 of the 46 galaxies in our primary sample, we obtain H$\alpha$ luminosity measurements (37/43) or upper limits (6/43) from the literature (see Table \ref{table:data} for references). For each of these systems, we calculate the contamination to this H$\alpha$ luminosity from evolved stars. 
%
To do this, we consider three different sources of ionization: (i) type Ia supernova remnants, (ii) planetary nebulae, and (iii) photoionization by evolved stars. For the first two contributions, we utilize observations of M31, where the number of planetary nebulae and supernova remnants is well characterized. From \cite{bhattacharya19}, we estimate a total integrated luminosity from planetary nebulae of L$_{H\alpha,PNe} \sim 2\times10^{37}$ erg s$^{-1}$, assuming L$_{[\textrm{O}\textsc{iii}]}$/L$_{H\alpha}$ = 4 for planetary nebulae \citep{davis18}. We can similarly sum up the total H$\alpha$ luminosity from type Ia supernovae remnants for M31, based on the survey of \cite{lee14}, finding L$_{H\alpha,SNR} = 7\times10^{37}$ erg s$^{-1}$. Combining these two, we find a contamination of $\sim$10$^{38}$ erg s$^{-1}$ to the H$\alpha$ flux coming from evolved remnants for an M31-mass galaxy. Given the stellar mass of M31 \citep[$10^{11}$ M$_{\odot}$;][]{sick15} and the relation between H$\alpha$ luminosity and SFR from \cite{kennicutt98}, this corresponds to a contamination of 0.008 (M$_*$/10$^{12}$ M$_{\odot}$) M$_{\odot}$ yr$^{-1}$ from evolved remnants (supernova remnants and planetary nebulae). To estimate the contribution to photoionization from evolved stars, we follow the prescriptions in \cite{cidfernandes11}. Assuming a single-age population older than 10$^8$ yr, \cite{cidfernandes11} estimate an ionization rate of $q_{HI} = 10^{41}$ s$^{-1}$ M$_{\odot}^{-1}$, which can be converted to a total H$\alpha$ luminosity assuming L$_{H\alpha}$($t>10^8$ yr) = ($h\nu_{H\alpha}/f_{H\alpha}$) $q_{HI}$M$_*$, where f$_{H\alpha} = 2.206$ for case B recombination. Under the same assumptions as above, this corresponds to a contamination of 1.1 (M$_*$/10$^{12}$ M$_{\odot}$) M$_{\odot}$ yr$^{-1}$ from low-mass, evolved stars. Given that this is two orders of magnitude higher than the contamination from remnants, we disregard the latter and consider only the contamination from low-mass, evolved stars.

To estimate the contamination to the measured H$\alpha$ luminosity, we consider only the flux within the spectroscopic aperture. Lacking the requisite information about aperture sizes/locations for each of the observations in Table \ref{table:data}, we consider two separate aperture corrections. For H$\alpha$ luminosities acquired from long-slit spectra, we assume a 1.5$^{\prime\prime}$ wide and 30$^{\prime\prime}$ long slit, and consider NGC~5044 as an example. Properly centered on the galaxy, such an aperture would contain 2.5\% of the total K$^{\prime}$-band luminosity (a good proxy for stellar mass), based on data from the Two-Micron All-Sky Survey \citep[2MASS;][]{2mass}. Thus, the contamination to long-slit spectra is $\sim$0.0275 (M$_{*,tot}$/10$^{12}$ M$_{\odot}$) M$_{\odot}$ yr$^{-1}$. For H$\alpha$ luminosities derived from IFU data or narrow-band imaging, we assume a fairly typical 10$^{\prime\prime}$ $\times$ 10$^{\prime\prime}$ extraction area \citep[e.g.,][]{mcdonald10,hamer16}, which encloses 10\% of the total K$^{\prime}$-band luminosity in NGC~5044, corresponding to a contamination of $\sim$0.11(M$_{*,tot}$/10$^{12}$ M$_{\odot}$) M$_{\odot}$ yr$^{-1}$. Finally, we make the conservative assumption that 50\% of these ionizing photons ultimately find a hydrogen atom, corresponding to 0\% escape and intrinsic extinction of E(B-V)=0.05 mag, or 50\% escape and no intrinsic extinction. We apply the relevant correction to each of the measurements in Table \ref{table:data}, depending on whether the H$\alpha$ luminosity was measured via long slit, IFU, or narrow-band imaging, leading to 5 additional upper limits. These revised H$\alpha$ luminosities, with the contribution from evolved stellar populations removed, are quoted in Table \ref{table:data}.

For each galaxy in our sample, we marginalize over the uncertainty on the L$_{H\alpha}$--SFR scaling relation, the uncertainty in the galaxy stellar mass (leading to an uncertainty in the contamination from evolved stars), the uncertainty on the K$^{\prime}$-band aperture correction described above (factor of 2 to account for different galaxy concentrations/sizes), and over the uncertainty in intrinsic extinction \citep{rosa-gonzalez02}, to arrive at an estimate of the SFR and its uncertainty for each system.%
%

Additionally, for 36/46 galaxies in our primary sample, we obtain molecular gas masses (M$_{H_2}$) from the literature (see Table \ref{table:data} for references). This provides an independent assessment of the amount of cold gas, which is less likely to be contaminated by mass loss from evolved stars or their remnants. We assume a constant depletion time of $\left<t_{dep}\right> = 2\pm0.5$ Gyr, consistent with the longer depletion times found for massive, mostly-passive galaxies \citep{huang14,genzel15,tacconi18}. The SFR is then simply assumed to be M$_{H_2}$/$\left<t_{dep}\right>$. We find that this is in good agreement with the H$\alpha$ derived star formation rate, as we show in subsequent sections. As an additional check, we also compare both of these estimates of the SFR to the classical cooling rates derived in \cite{mcdonald18a}, finding that they both correlate well and represent $\sim$1\% of the classical cooling rate on average.

For the three rapidly-accreting systems in our sample (Phoenix, H1821+643, IRAS09104+4109), we utilize slightly different SFR indicators. We note that, with star formation rates exceeding 100\,M$_{\odot}$ yr$^{-1}$ in these three systems, we don't need to worry about issues like contamination from old stellar populations.
To probe the warm, ionized gas, we use the [O\,\textsc{ii}]$\lambda\lambda$3726,3729 doublet, due to the fact the H$\alpha$ has redshifted to the near-IR for these systems. For Phoenix, we take the extinction-corrected [O\,\textsc{ii}] flux from \cite{mcdonald14a}, while for H1821+643 and IRAS09104+4109 we utilize new narrow-band observations with the \emph{Hubble Space Telescope} to measure the extended [O\,\textsc{ii}] without contamination from the central QSO (Calzadilla \etal in prep). In all three cases, we convert from L$_{[OII]}$ to SFR following \cite{kewley04}. Phoenix also has an estimate of the total molecular gas from \cite{russell17}, which we utilize in this work. For H1821+643 and IRAS09104+4109, we instead use the far-IR-derived SFR in place of an M$_{H_2}$-derived SFR, as neither of these clusters have a reliable estimate of the total molecular gas mass.

\subsection{Inferring M$_*$ and M$_{BH}$ via L$_K$}

For each galaxy in Table \ref{table:data}, we acquire the total K$^{\prime}$-band luminosity from the Two Micron All-Sky Survey \citep[2MASS;][]{2mass}, via the NASA Extragalactic Database\footnote{\url{http://ned.ipac.caltech.edu/}}. This is converted to a stellar mass assuming M$_*$/L$_K$ = 0.8 for red galaxies, based on \cite{mcgaugh14}. This stellar mass is used to correct for contamination in the H$\alpha$-derived SFR (as mentioned in the previous section), and to look for trends between BHAR/SFR and stellar mass of the host galaxy.

Black hole masses (M$_{BH}$) are taken from \cite{russell13}. This work utilized dynamic measurements when available and K-band luminosities of the host galaxy for all other galaxies, based on the scaling relation and methodology from \cite{graham07}. These black hole masses are only used in \S6, where we consider the \emph{specific} accretion rate (sBHAR $\equiv$ BHAR/M$_{BH}$) of the AGN in our sample.


\begin{figure*}[tb]
\centering
\includegraphics[width=0.99\textwidth]{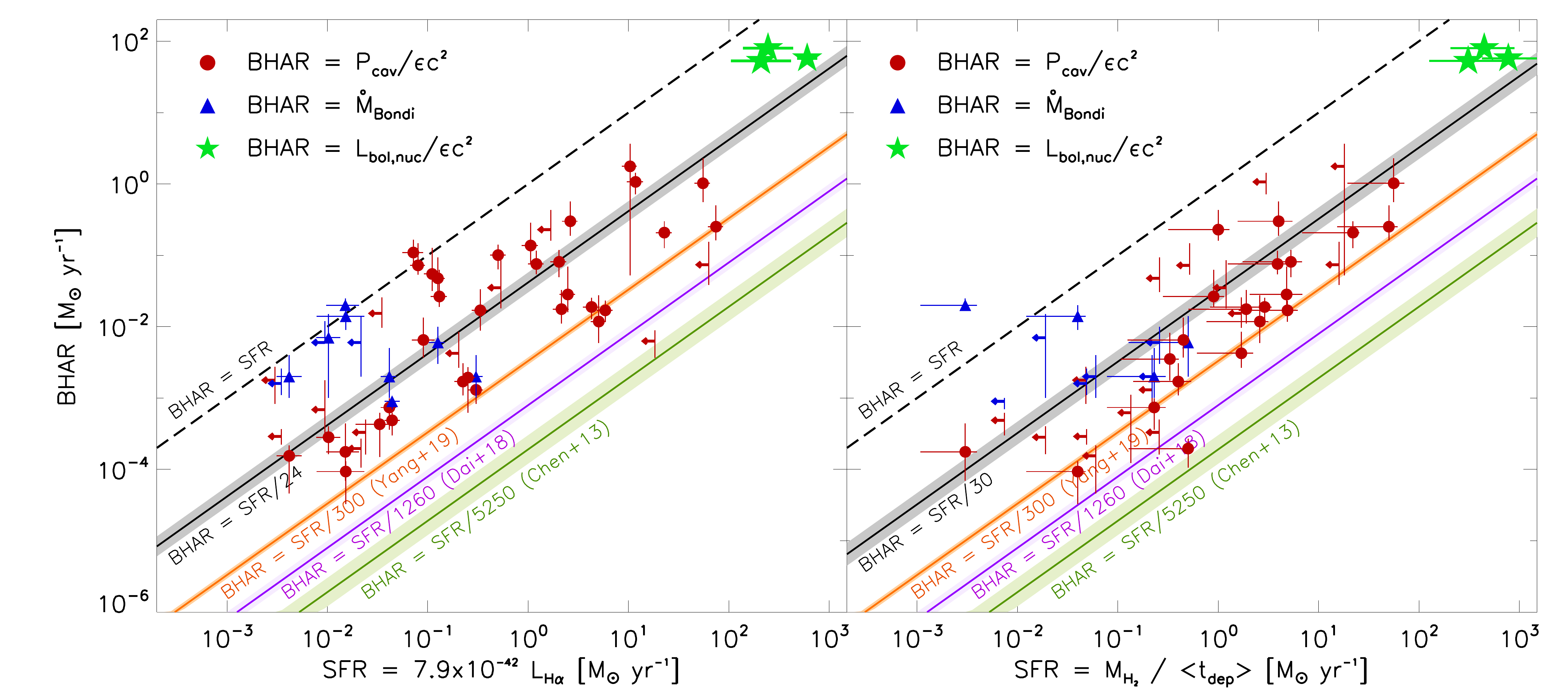}
\caption{Black hole accretion rate (BHAR) vs star formation rate (SFR) for our sample of giant elliptical galaxies. Star formation rates here are computed either via the H$\alpha$ emission line luminosity (left) or the molecular gas mass (right). Point color/type corresponds to how the accretion rate is constrained. For all data shown, we assume a ratio of energy output to matter accreted of $\epsilon_{acc}=0.1$. The solid black line and shaded gray region shows the best-fit BHAR/SFR ratio and the measured uncertainty on this ratio, respectively, incorporating non-detections by assuming an underlying log-normal distribution in the scatter.
For comparison, we show the relations and their uncertainty presented in \cite{chen13}, \cite{dai18}, and \cite{yang19b}.  Regardless of the method used to quantify the SFR or BHAR, giant elliptical galaxies at the centers of groups and clusters appear to be offset from the field population by roughly an order of magnitude, implying either a higher accretion efficiency (BHAR/SFR $\sim$ 4\%) or a higher energy output per unit mass accreted ($\epsilon\sim1$).}
\label{fig:bhacc}
\end{figure*}

\section{The SFR--BHAR Relation For Giant Elliptical Galaxies}

In Figure \ref{fig:bhacc} we show the relationship between black hole accretion rate (BHAR) and star formation rate (SFR) for our sample of giant elliptical galaxies. This figure includes accretion rates derived in three ways (based on jet power, the Bondi rate, and the bolometric X-ray luminosity) and star formation rate derived in two ways (H$\alpha$ luminosity, molecular gas mass). We utilize two distinct SFR indicators in an attempt to avoid bias and to ensure that our SFRs are reliable in the extremely low sSFR limit ($\lesssim$10$^{-4}$ Gyr$^{-1}$).
Supermassive black holes and their host galaxies in this sample span five orders of magnitude in both \.M/\.M$_{Edd}$ \citep{russell13} and specific star formation rate, yet span only one order of magnitude in BHAR/SFR. Regardless of the methodology used to estimate BHAR or SFR, we find that the typical BHAR/SFR ratio is $\sim$1/25. We compare this to estimates from the literature for a stacking analysis of spheroidal galaxies \citep[BHAR/SFR $\sim$ 1/300;][]{yang19b} and star-forming galaxies \citep[BHAR/SFR $\sim$ 1/3000;][]{chen13,dai18}. In general, we find elevated BHAR/SFR ratios for giant elliptical galaxies independent of whether they are rapidly accreting starburst QSOs (e.g., Phoenix, IRAS09104, H1821+643), or red-and-dead radio-loud galaxies. %
%
%
More specifically, we measure $\left<\log_{10}(\frac{BHAR}{SFR})\right> =  -1.38 \pm 0.14$ for H$\alpha$-derived SFRs, and $\left<\log_{10}(\frac{BHAR}{SFR})\right> =  -1.49 \pm 0.16$ for CO-derived SFRs. These measurements incorporate non-detections, assuming an underlying log-normal distribution of BHAR/SFR, and following the methodology of \cite{kelly07}, fitting only for the normalization in the BHAR--SFR relation. These measured values are considerably higher than the typical BHAR/SFR ratio quoted in the literature, as we will discuss in the following sections.
%
%

\begin{figure}[tb]
\centering
\includegraphics[width=0.47\textwidth]{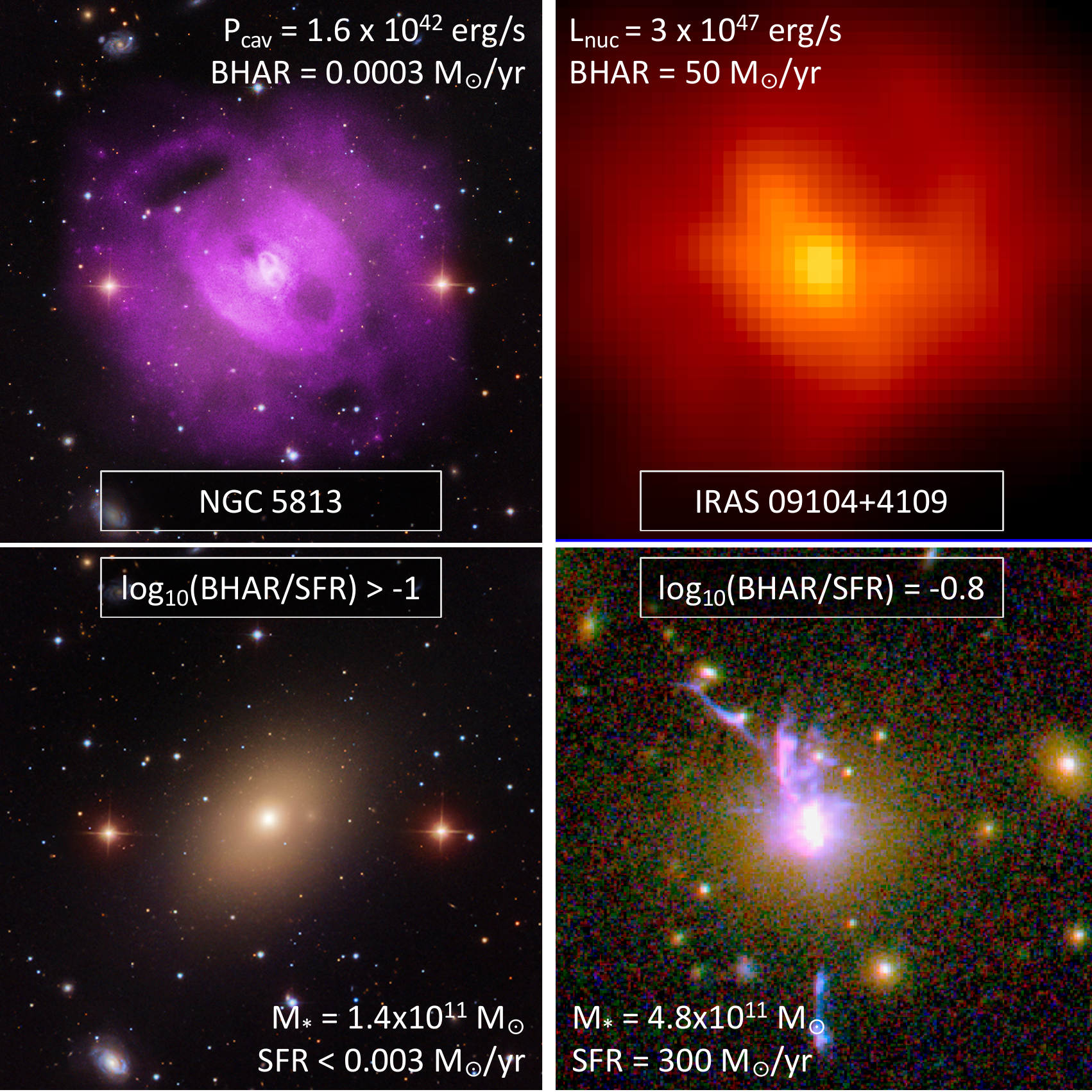}
\caption{Comparison of NGC5813 (left) to IRAS09104+4109 (right). These two galaxies, each living at the center of a massive galaxy cluster, have vastly different BHARs and SFRs. NGC5813 is a classic ``red and dead'' BCG, with powerful radio jets that have inflated multiple sets of bubbles in the hot ICM \citep{randall15}. IRAS09104+4109, on the other hand, is a starburst galaxy harboring a powerful type-II QSO at its center \citep{osullivan12}. Despite five orders of magnitude difference in the SFR and BHAR between these two central galaxies, they share very similar BHAR/SFR ratios of $\sim$0.1. While anecdotal, this figure highlights the huge dynamic range of this sample over which we measure relatively small variation in BHAR/SFR.}
\label{fig:images}
\end{figure}

To highlight the range in galaxy properties over which we observe a relatively small spread in BHAR/SFR ratios, we focus on two galaxies from this sample in Figure \ref{fig:images}. These two systems, NGC5813 and IRAS09104+4109, are both giant elliptical galaxies and are the central galaxies in clusters with masses of M$_{500} = 1.2 \times 10^{14}$ M$_{\odot}$ \citep{phipps19} and M$_{500} = 5.8 \times 10^{14}$ M$_{\odot}$ \citep{osullivan12}, respectively. Their BCG stellar masses are within a factor of $\sim$3 of each other, yet their star formation rates and black hole accretion rates are 5 orders of magnitude apart. Despite this huge difference, they have effectively the same BHAR/SFR ratio ($\sim$0.1) to within the measurement uncertainty. In both cases the power output of the AGN is very well constrained. In NGC5813, \emph{three} generations of bubbles inflated by radio jets are unambiguously detected, and are used in combination to produce a precise constraint on the time-averaged power output of the AGN \citep{randall15}. In IRAS09104+4109, the central AGN is a type-2 (heavily obscurred) QSO, with excellent constraints on the bolometric luminosity coming from a combined analysis of data from \emph{Chandra}, \emph{XMM-Newton}, \emph{SWIFT}, and \emph{Spitzer} \citep{vignali11}. These two systems highlight the diversity in cold gas supply that we observe within this sample of central galaxies, and suggest that the elevated BHAR/SFR ratios may be a universal property of giant ellipticals.

Before trying to interpret this high BHAR/SFR ratio, we consider whether such a high number makes sense in the context of cooling flows. The galaxies in our sample all sit at the center of a hot halo, with their central black hole regulating cooling on large ($\sim$100\,kpc) scales. This regulation comes from the mechanical jet power, which can be related to the accretion rate as:

\begin{equation}
\epsilon_{acc}\dot{M}_{BH}c^2 = P_{mech}
\end{equation}

\noindent{}where $\epsilon_{acc}$ relates the accretion rate to the power output, and is typically assumed to be 0.1. If we assume that cooling from the hot phase is well-regulated, we can write:

\begin{equation}
P_{mech} = \epsilon_{reg}L_{cool} = \epsilon_{reg}\frac{5kT\dot{M}_{cool}}{2\mu m_{\textrm{p}}}
\end{equation}

\noindent{}where $\epsilon_{reg}$ is how well cooling is being regulated, on average, and is observed to be $\sim$1 from studies of radio-mode feedback in groups and clusters \citep[e.g.,][]{rafferty06,mcnamara07,mcnamara12}. The cooling luminosity is the amount of energy radiated per particle as it cools from temperature $kT$. Finally, we can relate the cooling rate from the hot phase ($\dot{M}_{cool}$) to the star formation rate in the central galaxy as:

\begin{equation}
SFR = \epsilon_{cool}\dot{M}_{cool}
\end{equation}

\noindent{}where $\epsilon_{cool} \sim 0.01$ (1\% of the cooling gas actually makes it to the cold phase) is the basis for the cooling flow problem \citep[see recent summary by][]{mcdonald19}. Combining equations 1, 2, and 3 gives:


\begin{equation}
SFR = 45.2 \left(\frac{\epsilon_{cool}}{0.01}\right)\left(\frac{\epsilon_{acc}}{0.1}\right)\left(\frac{1.0}{\epsilon_{reg}}\right)\left(\frac{5~\textrm{keV}}{kT}\right)\dot{M}_{BH}
\label{eq:4}
\end{equation}

\noindent{}where 5\,keV is the median temperature of the hot halos in our sample, $\epsilon_{cool}$ is the ratio of SFR to classical cooling rate, $\epsilon_{acc}$ is the fraction of rest mass converted to energy in accretion/feedback process, and $\epsilon_{reg}$ is the ratio of the cooling luminosity to the energy output from the AGN. This relation demonstrates that, if cooling is balanced by mechanical feedback in these halos (which all evidence suggests is the case), we \emph{expect} a BHAR/SFR ratio of $\sim$1/45, rather than the more commonly cited values of $\sim$1/400 in the literature. While this is a useful exercise to understand the normalization of the BHAR--SFR relation in giant elliptical galaxies, it is only illustrative for the most massive, central galaxies, where the star formation and black hole accretion are both fueled by cooling of the hot halo, which is mostly held in check by AGN feedback. One would not expect this to be the case for lower-mass galaxies, where a larger fraction of the baryons are in stars that are forming primarily out of cold gas in a disk.

\begin{figure}[tb]
\centering
\includegraphics[width=0.47\textwidth]{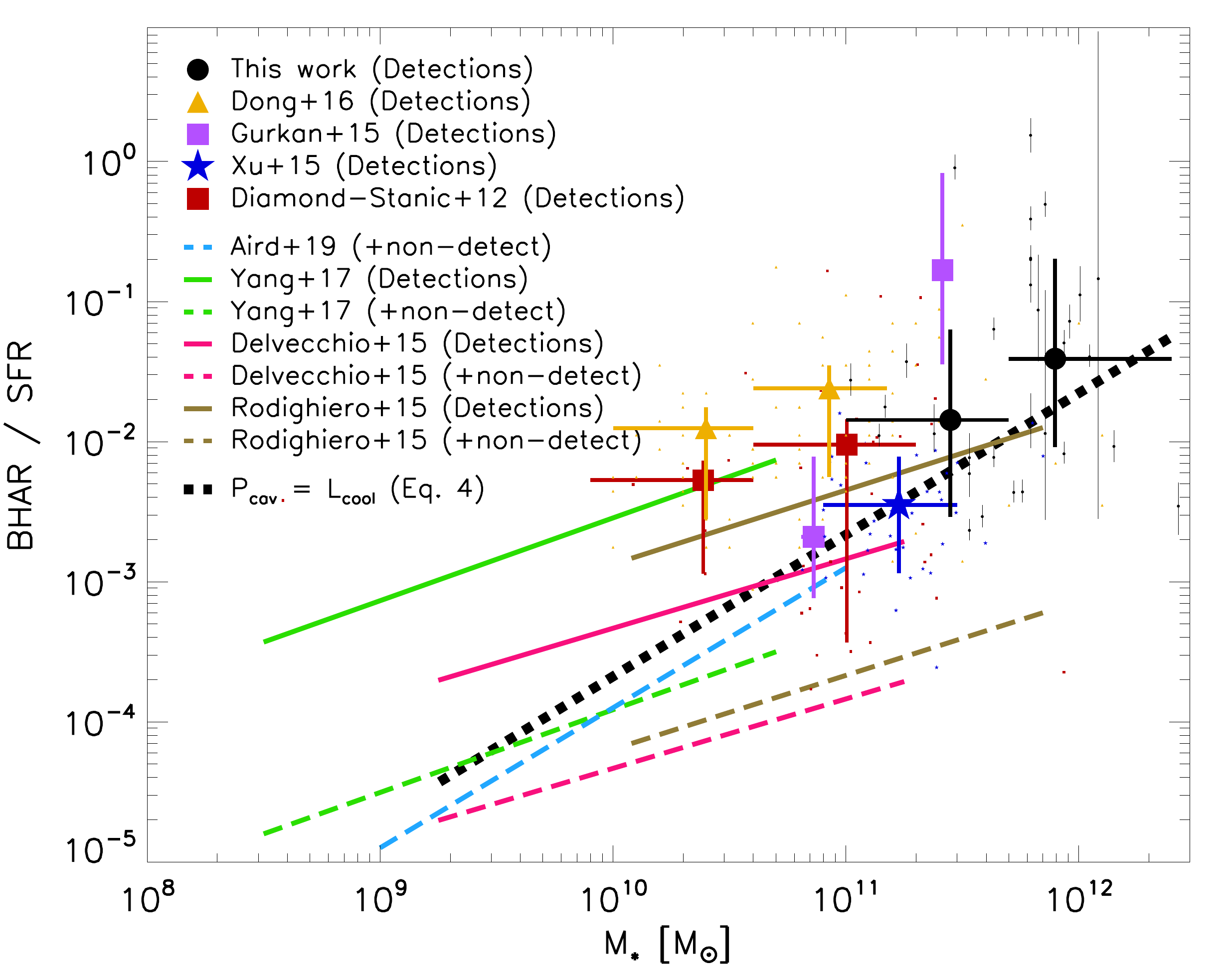}
\caption{Ratio of the black hole accretion rate (BHAR) to star formation rate (SFR) for galaxies from a variety of surveys. Data points are binned averages from this work (black) as well as the literature. Solid lines are best-fit relations from published works, including only detections (solid colored lines) and including non-detections (dashed colored lines). The dotted black line shows the relationship from Equation \ref{eq:4}, where we convert from halo temperature to stellar mass via \cite{anderson15}. This figure demonstrates that more massive galaxies have higher BHAR/SFR ratios in general and that, at the massive end, this is consistent with a self-regulating feedback loop where cooling is suppressed by AGN feedback.
}
\label{fig:mstar}
\end{figure}

Given that the galaxies in our sample are among the most massive in the Universe, it makes sense to consider whether the BHAR/SFR ratio has a mass dependence. In Figure \ref{fig:mstar} we show the average BHAR/SFR ratio measured by a variety of studies, as a function of stellar mass. For studies that incorporate non-detections, we attempt to scale out the contribution from stacking, considering only the detected AGN. We show, for comparison, the relation implied by Equation \ref{eq:4}, coupled with the relationship between central galaxy mass and halo temperature from \cite{anderson15}, which is derived from \cite{sun09a}. This relation, which suggests that BHAR/SFR $\propto$ M$_*$, describes the data well at the highest masses, but does not reproduce the shallow slope observed at lower masses, where the BHAR/SFR appears to scale more like M$_*^{2/3}$. We caution that there are considerable differences in sample selection and analysis that we will discuss in more detail in subsequent sections.

There are three simple interpretations of the high BHAR/SFR ratio measured in these massive, central galaxies. The black holes may be accreting more rapidly in these systems at fixed star formation rate (high BHAR), stars may be forming less efficiently at fixed black hole accretion rate (low SFR), or black holes may be converting a higher fraction of the accreted matter into energy than in lower-mass galaxies (high $\epsilon$). We will discuss each of these scenarios in more detail in \S5, attempting to determine which is responsible for the observed BHAR/SFR ratios.

\section{BHAR/SFR Ratios in the Literature}

In Figure \ref{fig:bhacc}, we showed that the BHAR/SFR ratio in giant elliptical galaxies is elevated compared to typical galaxies by factors of $\sim$10--100. However, with orders-of-magnitude variations in the quoted BHAR/SFR ratio within the literature, it is unclear whether this elevation is meaningful. Thus, we pause here to assess the state of the literature before attempting to interpret our own measurement of the BHAR/SFR ratio in giant ellipticals.

\begin{figure}[tb]
\centering
\includegraphics[width=0.47\textwidth]{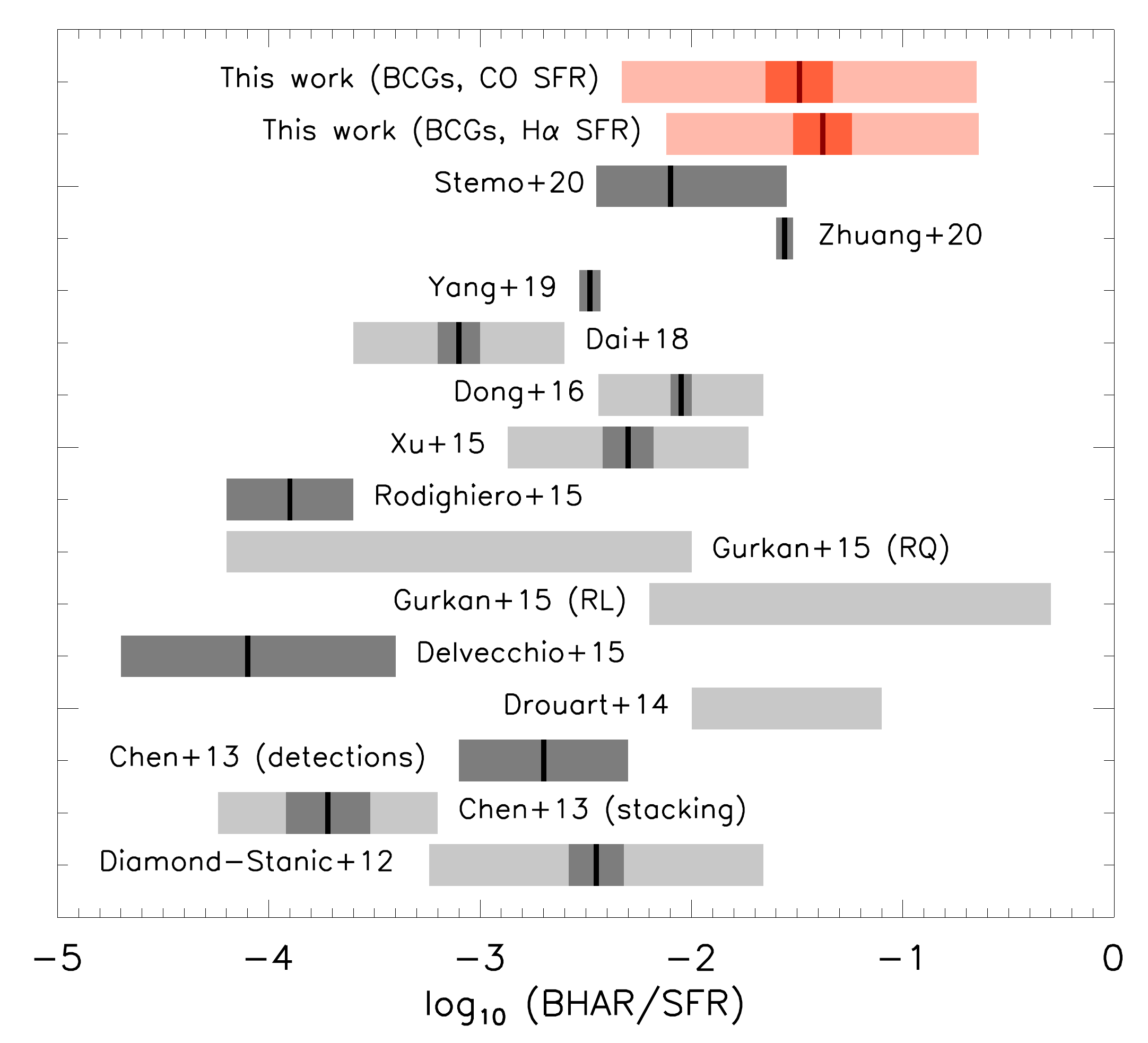}
\caption{The observed range in the BHAR/SFR ratio from the literature. Each row represents a different publication or subsample. Dark lines show the measured mean, dark bands show the uncertainty in the mean, and light bands show the scatter. All three measurements are not available for each sample. This figure highlights the huge range in BHAR/SFR quoted in the literature.}
\label{fig:literature}
\end{figure}

We present the results of our literature search in Figure \ref{fig:literature}, including data from \cite{diamond-stanic12}, \cite{chen13}, \cite{drouart14}, \cite{delvecchio15}, \cite{gurkan15}, \cite{rodighiero15}, \cite{xu15}, \cite{dong16}, \cite{dai18},  \cite{yang19b}, \cite{stemo20}, and \cite{zhuang20}. While this is not an exhaustive list of the published BHAR/SFR ratios, it represents all of the recent measurements that we are aware of at the time of writing, where the mean value of BHAR/SFR was quoted in the text, easily read off from a figure,  or the data were made available allowing us to make the measurement ourselves. Wherever possible we quote the mean value of the BHAR/SFR ratio, the uncertainty on the mean, and the measured scatter. Figure \ref{fig:literature} shows a huge amount of scatter in the measurements, from $\left<\log_{10}(\textrm{BHAR/SFR})\right> = -4.1$ \citep{delvecchio15} to $\left<\log_{10}(\textrm{BHAR/SFR})\right> \sim -1.5$ \citep{drouart14}, which reflects the wide variety in samples and methodology. 
For example, \cite{diamond-stanic12}, \cite{drouart14}, \cite{gurkan15},  \cite{xu15}, and \cite{dai18} consider only detected AGN, leading to systematically higher measurements of BHAR/SFR, while \cite{chen13}, \cite{delvecchio15}, \cite{rodighiero15}, and \cite{yang19b} all include non-detected AGN via stacking analyses, leading to systematically lower measurements of the BHAR/SFR ratio. \cite{chen13}, \cite{delvecchio15}, \cite{rodighiero15}, and \cite{dong16} focus on rapidly star-forming galaxies, while the samples of \cite{xu15} and \cite{yang19b} contain galaxies closer to the star-forming main sequence. Finally, the methodology used to compute the BHAR range from radio power \citep{gurkan15, drouart14}, to optical line luminosity \citep{diamond-stanic12, xu15}, to X-ray luminosity \citep{chen13, delvecchio15, rodighiero15, yang19b}. Given the variety of galaxy/AGN selection and methodology employed to measure the BHAR/SFR, it is perhaps unsurprising that there is so much scatter in the literature.

\begin{figure}[tb]
\centering
\includegraphics[width=0.49\textwidth]{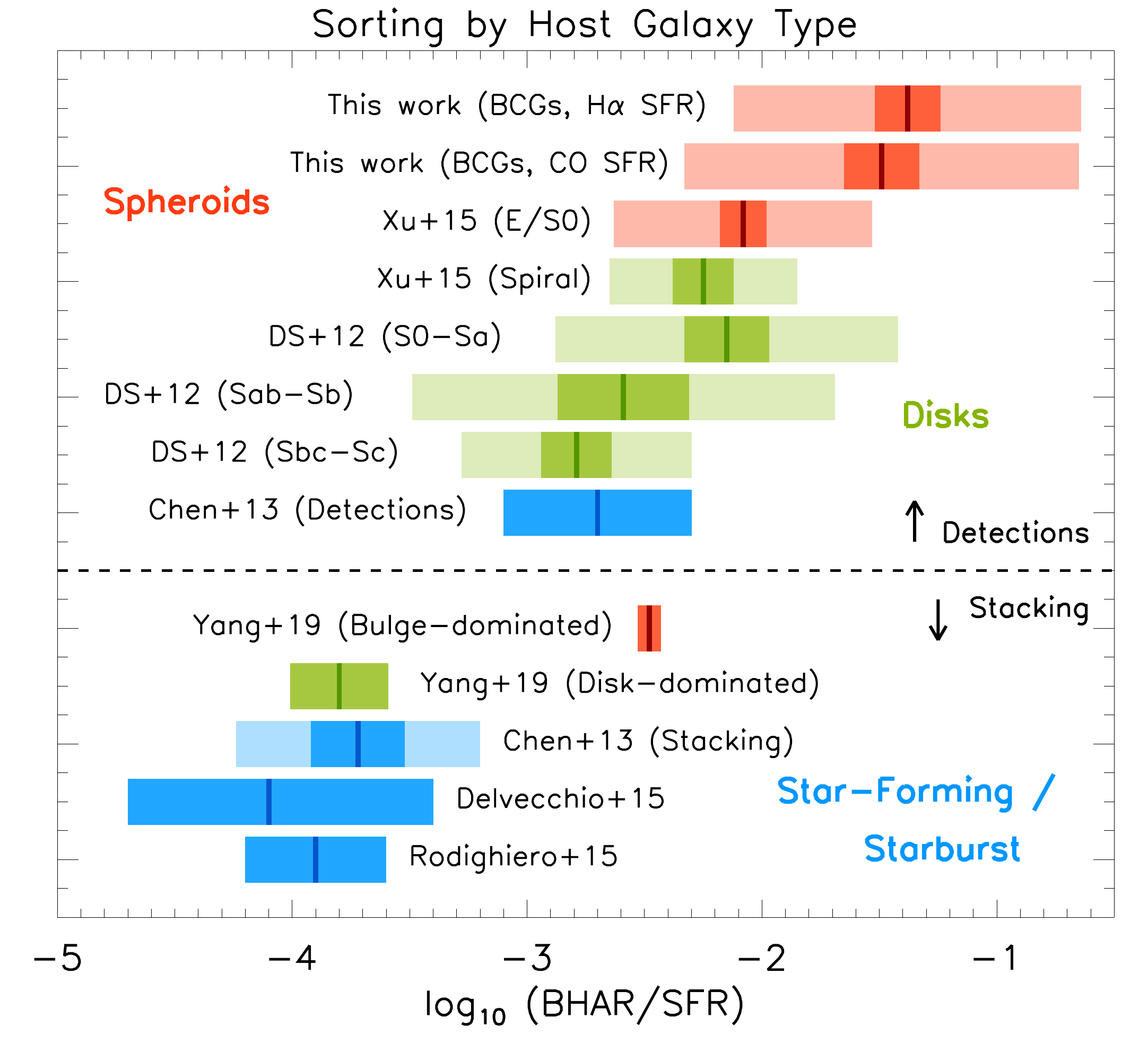}
\caption{Similar to Figure \ref{fig:literature}, but we have now grouped publications by galaxy type wherever possible. Publications with no information about host galaxy type have been excluded, including those that select blindly on AGN. Where the information is available, we show BHAR/SFR for subsamples by morphological type. Studies focusing on spheroidal galaxies are shown in red, disk galaxies in green, and star-forming or starburst galaxies in blue. We have also separated studies for which non-detections are included from those which focus only on detections -- the average BHAR/SFR are typically different by an order of magnitude for these two cases. This figure demonstrates that the measured BHAR/SFR ratio is strongly dependent on host galaxy morphology, with pure spheroids having roughly an order of magnitude higher BHAR/SFR than disk-dominated galaxies.}
\label{fig:literature_morph}
\end{figure}

In an attempt to understand the wide range of published BHAR/SFR ratios, we group publications by galaxy type and by the methodology used to constrain the BHAR. In Figure \ref{fig:literature_morph}, we consider only the subset of publications for which we have information about the host galaxy, either in the publication itself or via NED. Further, we separate the sample of literature estimates into those that incorporate non-detections and those that do not, recognizing that these subsamples will have disparate measurements of the average BHAR/SFR ratio. When sorting by galaxy type, we see a clear trend from star-forming disk galaxies (low BHAR/SFR ratio) to more passive, spheroidal galaxies (high BHAR/SFR ratio). This trend is perhaps most obvious in \cite{yang19b}, where they consider separately ``bulge-dominated'' and ``comparison'' (not bulge dominated) subsamples, finding a significant difference in the measured BHAR/SFR ratio with the same methodology. The trend is also apparent when subdividing the sample of nearby Seyfert galaxies published by \cite{diamond-stanic12} by morphological type, with the measured BHAR/SFR ratio being a factor of $\sim$5 times higher for S0-Sa galaxies than for Sbc-Sc galaxies. This figure makes clear that there is some correspondence between the properties of the host galaxy and the measured BHAR/SFR ratio, but whether that has to do with the stellar populations (late-type galaxies are more star-forming than early-type) or the galaxy morphology (late-type galaxies have disks, spheroids do not) remains an open question that we will return to later in the discussion.

\begin{figure}[tb]
\centering
\includegraphics[width=0.49\textwidth]{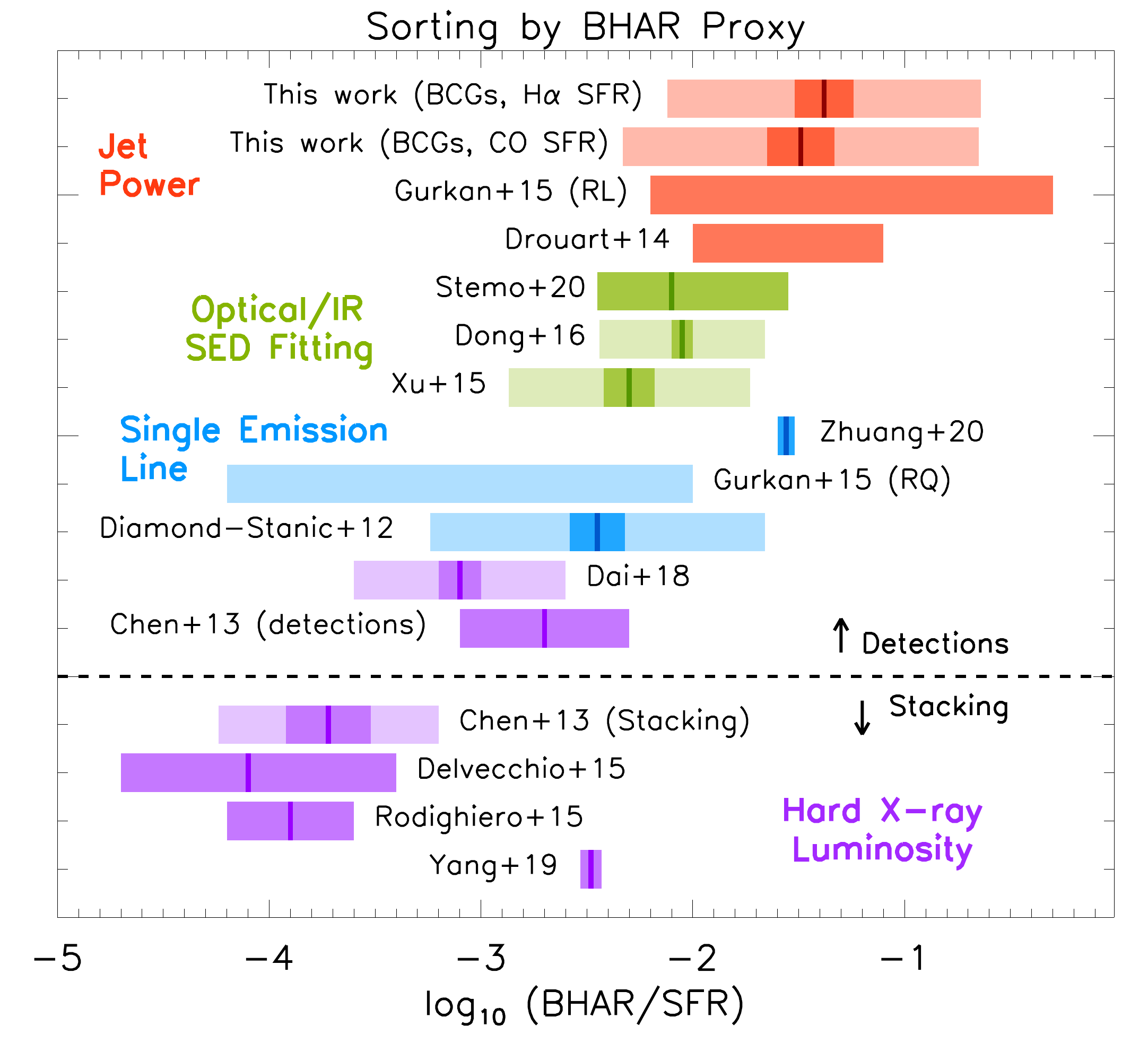}
\caption{Similar to Figures \ref{fig:literature} and \ref{fig:literature_morph}, but we have now grouped publications by AGN type. We have separated publications into radio-mode AGN (red), QSOs (green), optically-selected AGN (blue), and X-ray-selected AGN (purple). In general, radio-mode AGN tend to have higher BHAR/SFR than X-ray-bright AGN, with optically-selected AGN lying in between. The only outlier to this trend is the measurement by \cite{yang19b}, where the X-ray-bright AGN of spheroidal galaxies are found to be an order of magnitude higher than other X-ray-bright AGN. }
\label{fig:literature_agn}
\end{figure}

In Figure \ref{fig:literature_agn}, we subdivide the literature measurements by methodology used to constrain the BHAR. It is worth noting here that optical/IR proxies are biased towards high accretion rate (high $L/L_{Edd}$) systems, due to the dilution of the signal from the host galaxy  \citep[e.g.,][]{padovani17}. Given that the bulk of the literature sources that we compare to utilize X-ray- and radio-selected AGN, we are not concerned that such a bias is driving our results. 
 In general, Figure \ref{fig:literature_agn} shows that BHAR/SFR ratios estimated from the jet power are systematically higher than those measured from any other proxy, while X-ray luminosity tends to produce BHAR/SFR ratios that are systematically lower. This is somewhat counter-intuitive for two reasons. First, X-ray luminosity is highly variable, so we would expect to preferentially detect AGN where the X-ray luminosity is temporarily high, which ought to bias BHAR/SFR high. Second, X-ray-luminous AGN tend to be accreting closer to the Eddington rate, while radio-loud AGN are typically accreting at $\dot{M}/\dot{M}_{Edd} < 10^{-2}$ \citep[e.g.,][]{russell13}. For radio-loud galaxies to have both low Eddington ratios \emph{and} higher-than-average BHAR/SFR ratios, they would need to also have exceptionally low specific star formation rates, which is indeed the case. In general, we find much less scatter in the published BHAR/SFR ratios for AGN of a given type (X-ray, optical, radio). We find the most scatter when single emission lines (e.g., [O\,\textsc{iii}], [O\,\textsc{iv}]) are used, which makes sense since these require the largest and most uncertain bolometric corrections \citep[factors of 600--2500;][]{diamond-stanic12,zhuang20}. 

In summary, the published values of the BHAR/SFR ratio exhibit $\sim$3 orders of magnitude in scatter. Much of this scatter can be attributed to how non-detections are handled, with samples focusing on detected AGN finding systematically higher BHAR/SFRs than those that include non-detections. The remaining $\sim$1.5 orders of magnitude in scatter appears related to sample selection, with disk galaxies having considerably lower BHAR/SFR ratios than spheroidal galaxies, and X-ray AGN having considerably lower BHAR/SFR ratios than radio AGN. In the following sections, we attempt to disentangle these correlations, and determine the primary driver of scatter in the BHAR/SFR ratio and the origin of the high values observed in giant elliptical galaxies.

\section{Interpreting the High BHAR/SFR Ratios in Giant Elliptical Galaxies}

The high BHAR/SFR ratios that we measure in giant elliptical galaxies could be due to lower-than-average star formation rates, higher-than-average BH accretion rates, or a higher accretion efficiency ($\epsilon$) for radio power than for accretion luminosity. Below, we investigate each of these possibilities on an individual basis.

\subsection{Suppressed Star Formation}

\begin{figure}[tb]
\centering
\includegraphics[width=0.47\textwidth]{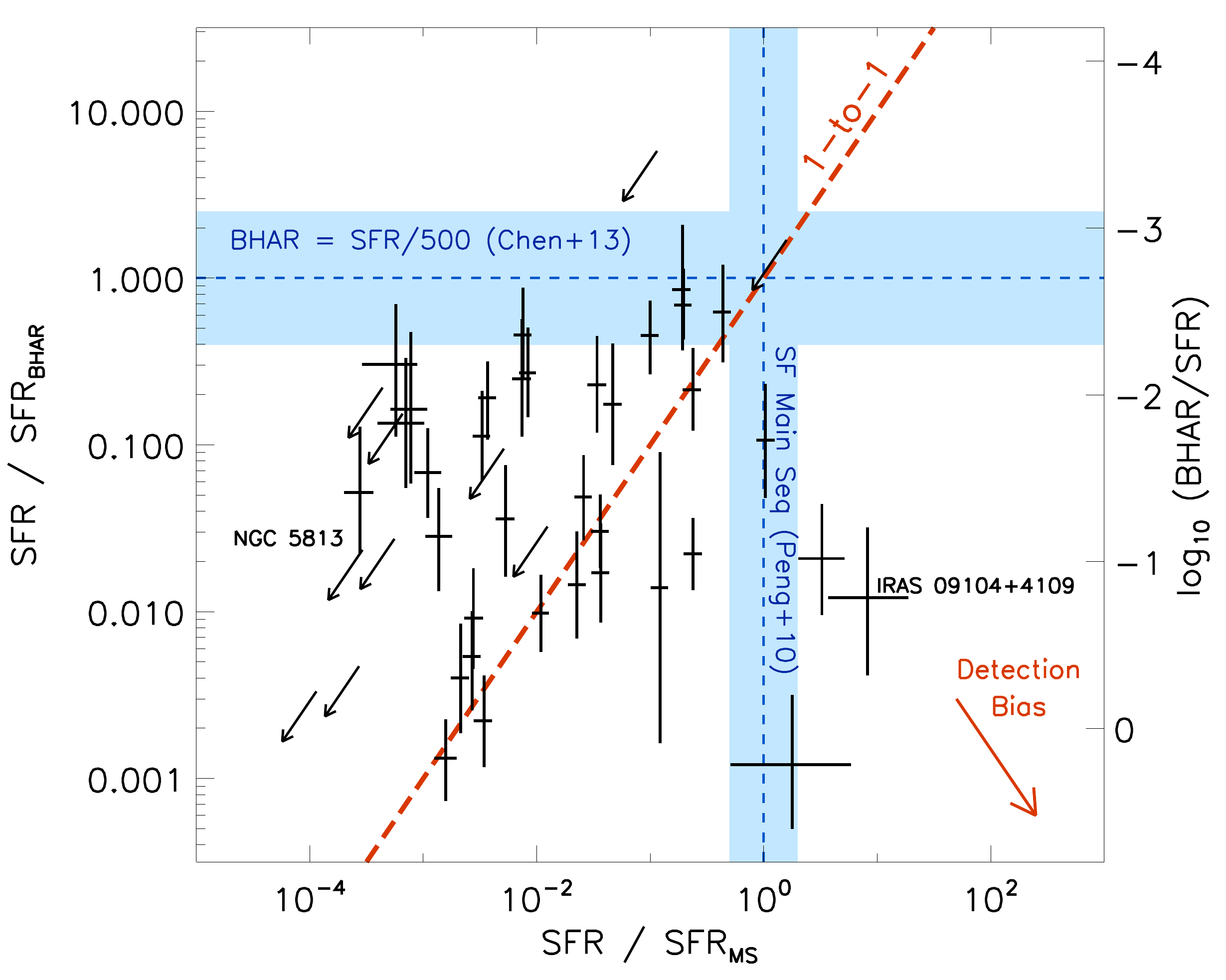}
\caption{Here we plot the offset from a canonical BHAR--SFR relation for detected AGN \citep{chen13} versus the offset from the main sequence of star formation \citep{peng10}. If the high BHAR/SFR ratios that we observed in our sample of BCGs is driven by them being more highly quenched, we would expect these two offsets to correlate. Instead, we see no evidence for a correlation (Pearson $r = 0.14$). We highlight the direction of selection bias  (towards bright AGN and star-forming galaxies), which does not appear to be masking any underlying correlation. Again, we highlight NGC5813 and IRAS09104+4109, which have very similar BHAR/SFR despite orders of magnitude difference in their position on the star-forming main sequence (see also Figure \ref{fig:images}).}
\label{fig:mainseq_justus}
\end{figure}

Giant elliptical galaxies are amongst the most quenched galaxies in the Universe, with star formation rates orders of magnitude lower than one would predict based on the amount of available fuel in the hot halo. As such, an obvious explanation for the high BHAR/SFR ratio that we observe is that black hole feedback is more effectively preventing stars from forming than in typical galaxies. We investigate whether this is the case in Figure \ref{fig:mainseq_justus}. We consider a baseline BHAR = SFR/500 relation for detected AGN \citep{chen13}, and ask whether deviations from this relation correlate with deviations from the main sequence of star formation \citep{peng10} -- if the high BHAR/SFR ratios that we observe are due exclusively to suppressed star formation, then the deviations between these two relations should correlate 1-to-1. We find no evidence that this is the case. In general, giant elliptical galaxies fall below the star forming main sequence and have elevated BHAR/SFR ratios, but these offsets are not correlated (Pearson $r = 0.08$). 
While our sample is incomplete, we would expect selection biases to drive us towards high SFR and high BHAR. Instead, as the observed star formation rates get further from the main sequence, the BHAR/SFR ratio remains roughly constant, despite the fact that higher BHAR/SFR systems ought to be easier to detect (larger cavities, brighter X-ray point sources). Again, we highlight NGC5813 and IRAS09104+4109 in this plot, which lie on the same BHAR--SFR relation, yet have $\sim$4 orders of magnitude difference in their specific star formation rate.

\begin{figure}[tb]
\centering
\includegraphics[width=0.47\textwidth]{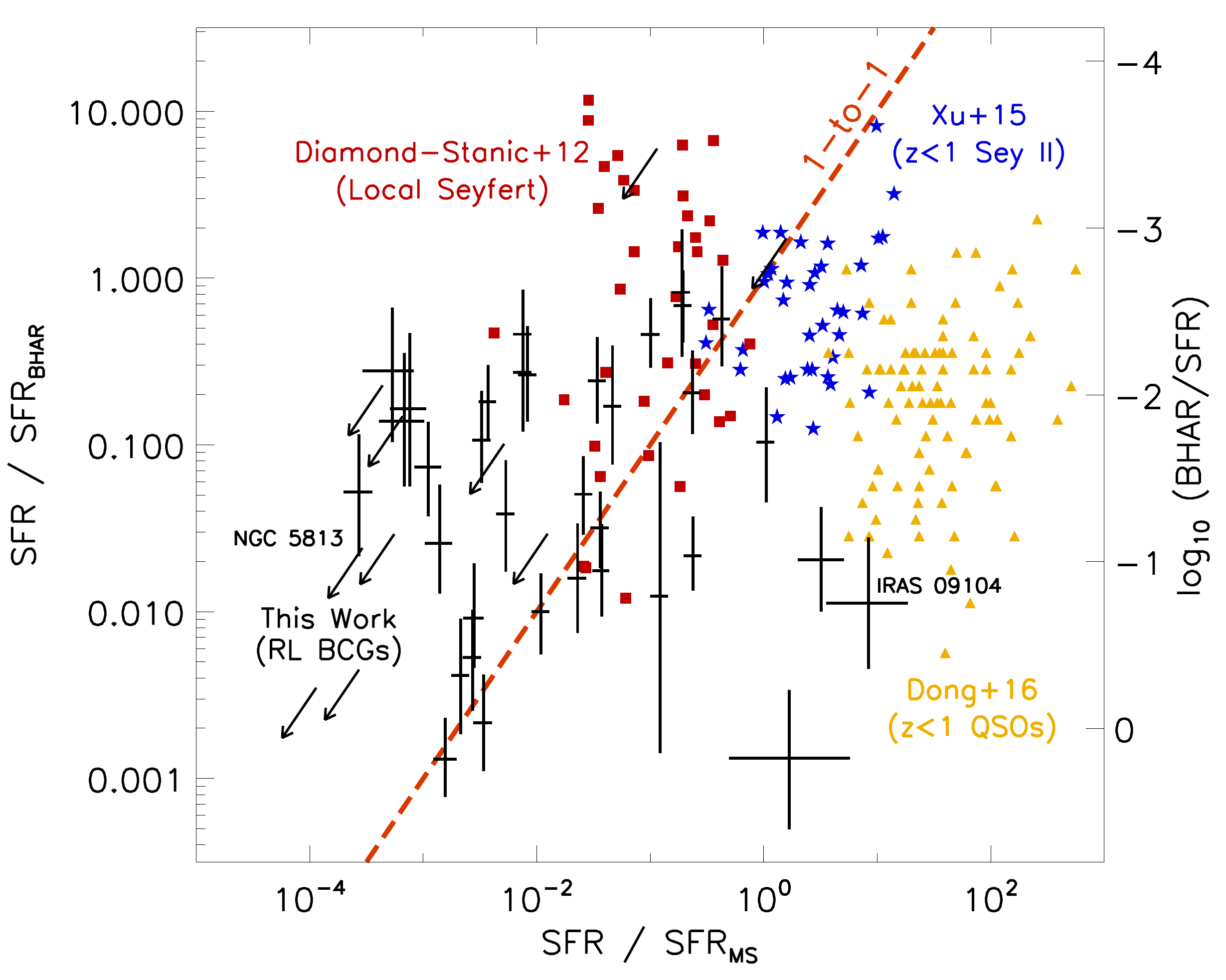}
\caption{Similar to Figure \ref{fig:mainseq_justus}, but now showing three other surveys as well, where the appropriate data are available \citep{diamond-stanic12, xu15, dong16}. This figure demonstrates that the offset from the star-forming main sequence does not appear to correlate with the BHAR/SFR ratio across four very different samples.}
\label{fig:mainseq}
\end{figure}

In Figure \ref{fig:mainseq} we consider additional systems from the literature, where individual measurements of the SFR, BHAR, and stellar mass are available \citep{diamond-stanic12, xu15, dong16}. Again, we see no correlation when we consider galaxies ranging from extremely passive (4 orders of magnitude below star forming main sequence) to starbursting (2 orders of magnitude above star forming main sequence). In general, the higher BHAR/SFR ratios are observed in galaxies with a wide variety of stellar populations (passive to starburst), while the lowest ratios are observed primarily in galaxies near the main sequence of star formation. Individually, or as an ensemble, these additional data from the literature do not exhibit a correlation between the distance from the star forming main sequence and the BHAR/SFR ratio.

Given that we do not find any link between how passive a galaxy is (distance from SF main sequence) and the BHAR/SFR ratio, we infer that the high BHAR/SFR ratios in giant ellipticals are not simply due to these galaxies being significantly more passive than the typical field galaxy. As such, we investigate alternative explanations below.

\subsection{Enhanced Black Hole Accretion}


Figure \ref{fig:literature_morph} highlights a strong correlation between galaxy morphology and the mean BHAR/SFR ratio. Galaxies that are more spheroidal tend to have higher BHAR/SFR, implying that a higher fraction of the available cold gas makes it in to the central black hole before forming stars. This trend is observed for individually-detected AGN \citep{diamond-stanic12,chen13,xu15} and, more importantly, for analyses that include non-detections via stacking \citep{chen13, delvecchio15, rodighiero15,yang19b}. In particular, \cite{yang19b} find an order of magnitude difference in the BHAR/SFR ratio when they consider bulge-dominated and disk-dominated galaxies separately, keeping all other aspects of their analysis the same. Given that the high BHAR/SFR ratio does not appear to be driven by the SFR, we consider whether it is plausible that the BHAR may actually be enhanced in spheroidal galaxies.

\begin{figure}[tb]
\centering
\includegraphics[width=0.47\textwidth]{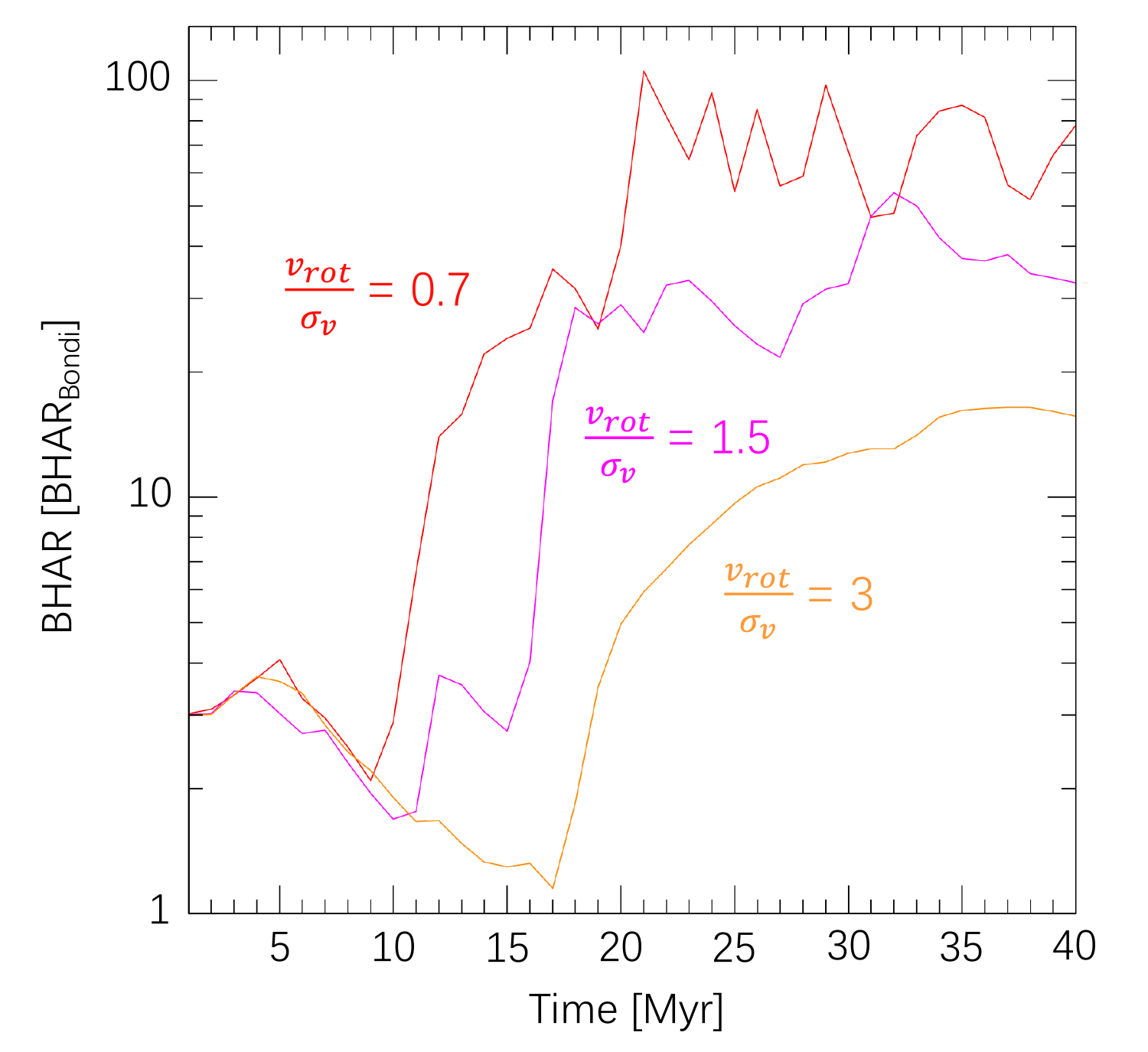}
\caption{Adapted from \cite{gaspari15b}, who use 3D hydrodynamic simulations to probe the impact of rotation on hot and cold accretion flows in a massive galaxy. This plot shows the evolution of the black hole accretion rate, normalized to the Bondi rate, as a function of time and as a function of the dynamics of the hot atmosphere. For non-rotating hot atmospheres ($v_{rot}/\sigma_v < 1$, i.e., turbulent Taylor numbers below unity), chaotic collisions between cold filaments/clouds that are condensing out of the hot phase promote the cancellation of angular momentum, leading to high accretion rates. As $v_{rot}/\sigma_v$ increases, the accretion flow shifts from turbulence-driven (linked to extended filaments and boosted accretion) to rotationally-driven (tied to a coherent disk and suppressed accretion).}
\label{fig:gaspari}
\end{figure}

\cite{gaspari15b} investigate whether black hole accretion is affected by large-scale rotation in the hot atmosphere. Using 3D hydrodynamic simulations, they simulate a massive galaxy embedded in a hot halo, including feedback from a central supermassive black hole and turbulence in the hot gas. As cool clouds condense from the hot gas, recurrent collisions and tidal forces between clouds, filaments and the central clumpy torus promote angular momentum cancellation, boosting accretion in a process known as ``chaotic cold accretion'' \citep[CCA;][]{gaspari13b}. As rotation is added to the hot halo, the accretion rate slows, as the accretion flow shifts from turbulence-driven to rotationally-driven (Figure \ref{fig:gaspari}). \cite{gaspari15b} find that the steady-state accretion rate can drop by a factor of $\sim$10 by dialing up the rotation, with the accretion mode also qualitatively changing from a clumpy rain to a coherent disk. The latter suppresses accretion onto the central supermassive black hole due to the high angular momentum of the gas, but does not hinder the formation of stars which form efficiently in large-scale disks. Furthermore, connecting the BHAR to the dynamics of the hot halo would naturally reproduce the observed correlations between black hole mass and the properties of the host galaxy's hot halo \citep[e.g.,][]{gaspari19}.

In general, this picture of a transitioning accretion mode from disk-dominated to spheroidal galaxies is consistent with observations. Most well-known AGN show evidence of a rotating disk of cool gas near the center -- indeed, this is often how the black hole mass is calculated. On the contrary, rotating disks of cool gas are rarely seen in giant ellipticals, with a few notable exceptions \citep[e.g., Hydra A;][]{rose20}. Instead, observations of these massive galaxies have found radial filaments of cool gas in emission \citep{johnstone87,crawford99,conselice01,edwards09,mcdonald10,mcdonald11a,hamer16}, and discrete, cold clumps in the vicinity of the central black hole in absorption \citep{tremblay16, rose19, rose20, schellenberger20}. This cool gas is thought to have condensed out of the slow-moving hot phase, perhaps aided by uplift from the radio jets, which should result in minimal angular momentum. While this gas will ultimately form into a disk near the center of the galaxy, the journey to smaller radii is shortened by the lessened angular momentum, which translates to less time to form stars and overall lower BHAR/SFR ratios.


It seems plausible that the BHAR/SFR ratio may be high in giant elliptical galaxies because the central supermassive black holes are able to accrete a larger fraction of the large-scale cold gas supply in dispersion-dominated systems than in rotation-dominated systems. This is consistent with simulations (in particular, those studying the above CCA process; Figure \ref{fig:gaspari}), and makes qualitative sense based on the different angular momentum configurations of spheroids versus disks, which should dictate how much gas can arrive at the center.

\subsection{Enhanced Feedback Efficiency and \\Black Hole Spin}

The trend in Figure \ref{fig:literature_morph} may indicate that spheroidal systems are experiencing unusually high jet power because their nuclear black holes are accreting more efficiently.  However, this interpretation rests on the assumption that the conversion efficiency between accretion rate and power, $\epsilon$, is constant over all systems.  This need not be so.  The value of $\epsilon$ depends primarily on the spin of the nuclear black hole.  As the spin parameter approaches unity, the radius of the innermost stable circular orbit contracts from $6R_g$ for a hole with zero angular momentum to $R_g$ for a maximally spinning hole with spin parameter $j=0.998$.  This contraction enables the accreting matter to fall deeper into the potential well yielding higher AGN power per gram of accreted matter,~$\epsilon$.


Most of the energy released in the systems studied here is in the form of jets, which typically form when the accretion rate slows to only a few percent of the Eddington accretion rate.  The most plausible mechanism for jet formation is the \cite{blandford77} process and its variants \citep[e.g.,][]{meier01,nemmen07}.  Rotational and gravitational binding energy are channelled into bi-directional jets mediated by strong magnetic fields twisting outward along the black hole's spin axis.  In these models, as the black hole's spin increases to its maximum rate, $\epsilon \rightarrow 0.42$. 

\begin{figure}[t]
\centering
\includegraphics[width=0.49\textwidth]{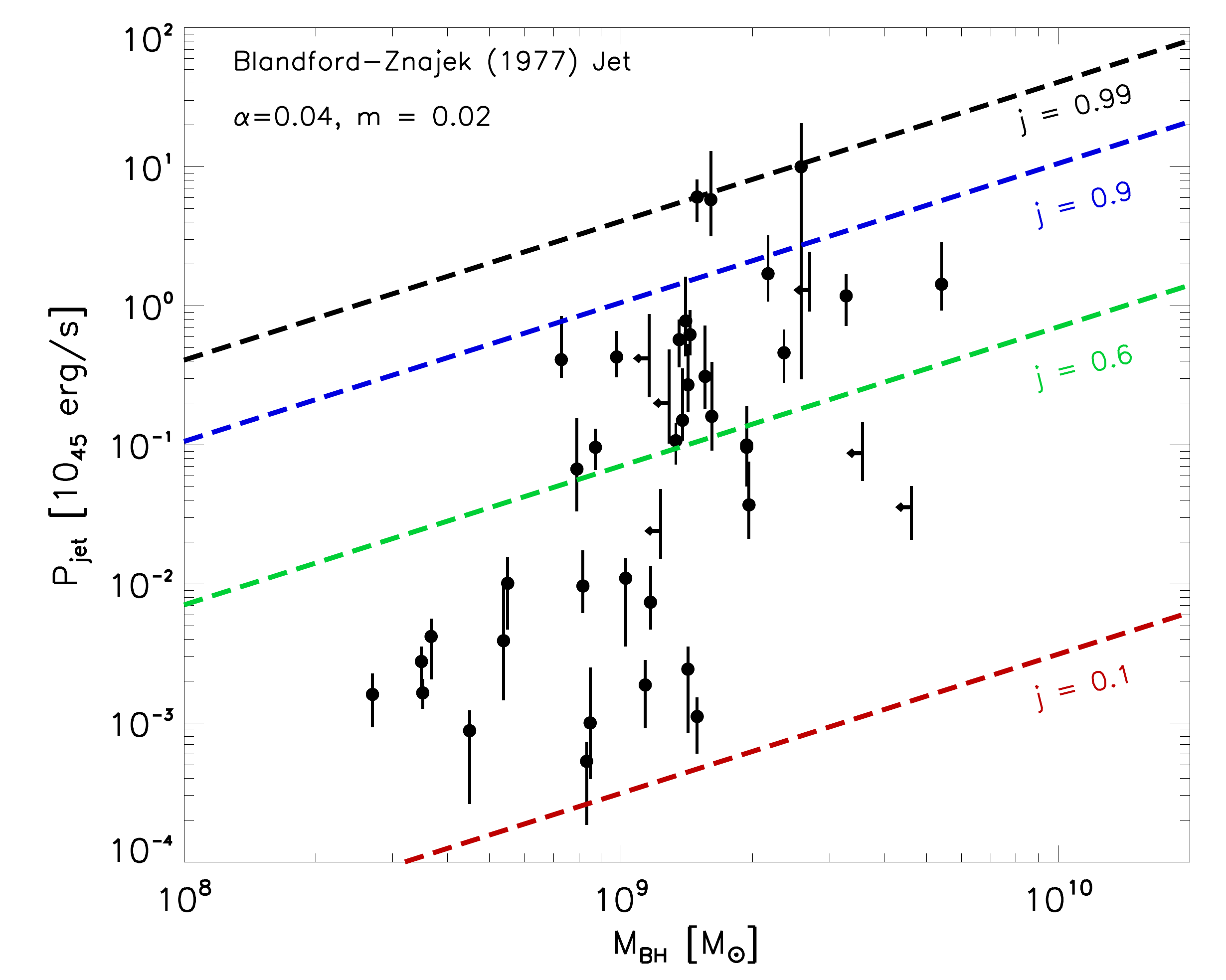}
\caption{Jet power as a function of black hole mass for central cluster and group galaxies from \cite{russell13}. Diagonal lines show models from \cite{nemmen07} with low viscosity ($\alpha=0.04$) and a constant accretion rate normalized to the Eddington rate ($m\equiv \dot{M}/\dot{M}_{Edd} = 0.02$). This figure demonstrates that, for a fixed accretion rate, the jet power can vary by four orders of magnitude, due to variations in spin. As such, our entire sample of massive galaxies is consistent with a single normalized accretion rate, under the assumption of a \cite{blandford77} jet model. This model was chosen as an extreme case to demonstrate a point -- we suspect that much of the scatter in P$_{jet}$ is, in fact, driven by variations in accretion rate, given the huge range in observed star formation rates in these systems.}
\label{fig:spin}
\end{figure}

This connection between jet power and spin is illustrated in Figure \ref{fig:spin}, where we show the degree to which jet power is enhanced by increasing angular momentum, $j$, of the hole. 
Following \cite{nemmen07} and \cite{mcnamara11}, we compute the jet power as a function of black hole mass for a  rotating black hole. We assume an accretion rate of $m\equiv \dot{M}/\dot{M}_{Edd}=0.02$ and disk viscosity ($\alpha = 0.04$) for illustration. This figure shows how sensitive the jet power is to the black hole spin, spanning 4 orders of magnitude at fixed accretion rate as the spin increases from from 0.1 to 0.99. We include data from \cite{russell13} in this figure, confirming that their jet powers can be achieved by assuming by a fixed, sub-Eddington accretion rate while varying spin. Choosing instead Nemmen's ``hybrid'' jet model with a higher, more realistic, viscosity \citep[$\alpha=0.3$;][]{nemmen07}, we find a two decade spread in P$_{jet}$ at fixed accretion rate.

To some extent, Figure \ref{fig:spin} suggests an even \emph{larger} BHAR/SFR ratio for jet-dominated AGN. For all four of the jet models presented in \cite{nemmen07}, they find $\epsilon<0.1$ for $j<0.9$. That is, for black holes spinning below the maximal rate, the implied BHAR/SFR would actually be \emph{higher} than what we quote, making the discrepancy with other publications worse. While there are too few measurements, some evidence indicates the most massive ($\gtrsim$10$^8$ M$_{\odot}$) black holes have intermediate spins (0.4--0.6), based on X-ray reflection spectroscopy \citep{reynolds14,reynolds19}. While these measurements are indicative, they are uncertain and cannot exclude the possibility that the most massive black holes in the universe are spinning maximally. In this case, the hybrid jet models of \cite{nemmen07} yield $\epsilon = 0.2-0.4$, implying that our estimates of the BHAR are high by factors of 2--4. Such a correction would bring our estimate of the BHAR/SFR ratio in line with those based on optical/IR SED fitting \citep[e.g.,][]{xu15,dong16,stemo20}. However, in order to have consistency between estimates of BHAR derived from jet power and those derived from hard X-ray luminosities, we require a more extreme value of $\epsilon=1$.

Simulations of accreting black holes with general relativity and magnetohydrodynamics have demonstrated that magnetic fields can impede accretion, magnetically arrest the disc, and drive powerful outflows. For low-spin black holes, these magnetically-arrested discs \citep[MAD;][]{narayan03} can drive jets with efficiencies of $\epsilon \sim 0.3$, while for maximally-spinning black holes the efficiency can be as high as $\epsilon \sim 1.4$ \citep{tchekhovskoy11}. This factor of $\sim$10 increase in $\epsilon$ above the canonical $\epsilon=0.1$ value could fully explain the factor of $\sim$10 difference that we observed in the BHAR/SFR ratio for radio galaxies compared to optical/X-ray AGN (Figure \ref{fig:literature_agn}).
This possibility is attractive, as it would naturally explain, using a unified model, the difference between accretion rates derived via jet powers and disk luminosities, as shown in Figure \ref{fig:literature_agn} and presented in \cite{gurkan15}.  
On the other hand, MAD requires some fine-tuning in the setup to induce such a high-efficiency mode (i.e., extremely large poloidal and coherent magnetic flux, as well as BH spins approaching unity). At variance, other GR-MHD simulations \citep[e.g.,][]{sadowski17} find that the horizon efficiency is stable around 4\% over five orders of magnitude in Eddington ratio and with varying physics. 
%

Despite some drawbacks, invoking spinning black holes is appealing due to the fact that the increased accretion efficiency is theoretically motivated and resolves much of the scatter in the observed BHAR/SFR ratio. 
However, there remains the issue of how these black holes achieve such high spins.
Since $z\sim 1$, these systems have grown primarily via dry mergers with ever-smaller satellites, which is not thought to yield high spin factors \citep{hughes03}. While \cite{volonteri07} find that elliptical galaxies tend to have central black holes with high spin, this is based on a picture where elliptical galaxies are formed exclusively via the merger of two gas-rich galaxies, following \cite{hopkins06}. On the other hand, the central galaxies in our sample have most likely grown via accretion of ever-smaller satellites over the past $\sim$10\,Gyr \citep{delucia07,lidman12}. \cite{volonteri05} found that, when black hole growth was connected to the hierarchical growth of galaxies, the most massive galaxies (M $= 10^{12}$ M$_{\odot}$) had spin distributions that are flatter, peaking around $a \sim 0.5$. This lower spin is due to these black holes growing primarily via mergers with smaller holes, rather than via rapid accretion. If the spin \emph{were} higher in these systems, the implied accretion rates based on the measured jet powers would then be \emph{lower}, exacerbating the already-large discrepancy between the growth rate from mergers and from gas accretion.

To summarize, while simulations predict that rapidly-spinning black holes may convert a larger fraction of the accreted mass into energy, there remain 
several challenges to this as an explanation for the high BHAR/SFR ratios that we observe in giant elliptical galaxies. It is unclear how these massive black holes could achieve such high spins, given that they (probably) grew most of their mass via mergers with smaller black holes over the past several Gyr. Further, to achieve efficiencies approaching unity is challenging and requires fine-tuning of the simulations.
Instead, we feel that the scatter in the BHAR/SFR ratio is more likely reflecting a higher BH accretion rate in spheroidal galaxies compared to disk galaxies, perhaps due to the angular momentum of the gas fueling both star formation and BH accretion as we discuss in \S5.2, with differences in accretion efficiency ($\epsilon$) playing a secondary role.

\section{Addressing Detection Bias and Consistency with the M$_{\textrm{BH}}$--M$_*$ Relation}


Correlations between the masses of supermassive black holes and their host galaxies have been well-studied for decades. Of these, the most well characterized are the relations between the black hole mass and spheroid luminosity and the black hole mass and spheroid velocity dispersion \citep[see review by][]{kormendy13}. While there is considerable scatter in the published relations \citep[see e.g.,][]{schutte19}, there is fairly broad consensus that black hole masses scale nearly linearly with spheroid masses, with \cite{kormendy13} finding M$_{BH}$ $\sim$ 0.005M$_{*, bulge}$ for ellipticals and classical bulges. This correlation is considerably weaker when the total stellar mass is considered in disk-dominated galaxies, as shown by \cite{reines15}. 

\begin{figure}[tb]
\centering
\includegraphics[width=0.47\textwidth]{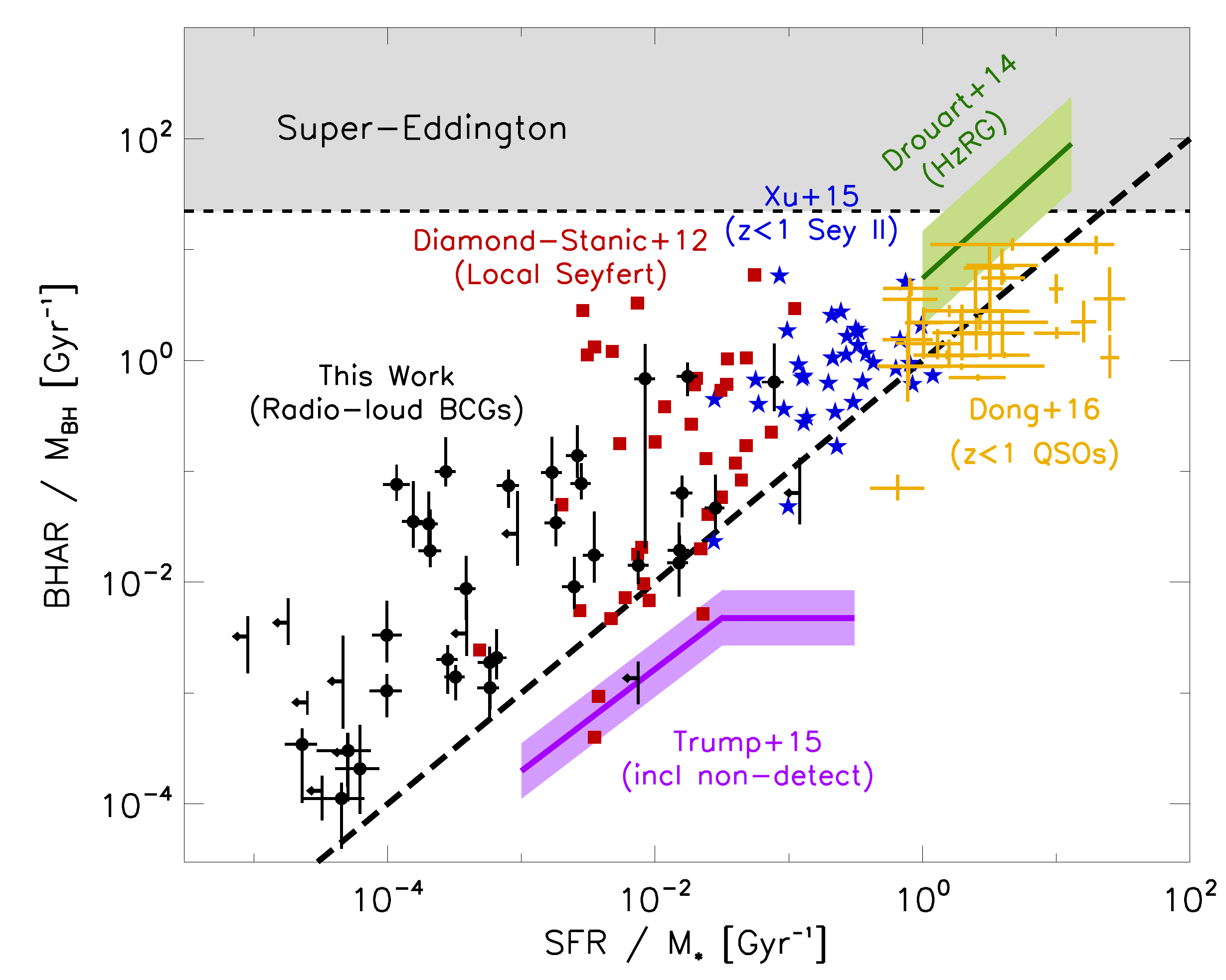}
\caption{Specific BHAR ($\textrm{sBHAR} \equiv \textrm{BHAR}/\textrm{M}_{\textrm{BH}}$) as a function of specific SFR  ($\textrm{sSFR} \equiv \textrm{SFR}/\textrm{M}_{\textrm{*}}$). We show data from this work, alongside literature values wherever available. The dashed black line shows the 1-to-1 relation -- systems above this line have black holes growing faster than the host galaxy, while those below the line have galaxies growing faster than their black hole. Given that all of the galaxies on this plot were selected on the presence of an AGN, it is unsurprising that they all lie above the 1-to-1 relation. Interestingly, there is very little scatter over 6 orders of magnitude in both sBHAR and sSFR, suggesting that the physical processes governing accretion and star formation are remaining relatively constant across all galaxies. As we discuss in the text, this correlation is predicted by the competing dependence on the bulge-to-total ratio between the M$_{\textrm{BH}}$--M$_{\textrm{spheroid}}$ and BHAR--SFR relations.
}
\label{fig:specific}
\end{figure}

Given that the aforementioned scaling relations are between the black hole mass and the \emph{spheroid} mass (rather than the total mass), the black hole accretion rate and the 
\emph{total} star formation rate do not necessarily need to scale the same way for consistency. The relation between black hole mass and spheroid mass can be rewritten as M$_{BH}$ $\sim$ 0.005$\cdot$(B/T)$\cdot$M$_*$, where now M$_*$ is the total stellar mass of the host galaxy and $B/T$ is the bulge-to-total ratio. If we assume that the BHAR/SFR ratio scales with B/T, as Figure \ref{fig:literature_morph} seems to imply, we expect that the \emph{specific} BHAR/SFR ratio should be a constant across all galaxies, since M$_{BH}$/M$_*$ scales with B/T as well. That is, the ratio of sBHAR $\equiv$ BHAR/M$_{BH}$ and sSFR $\equiv$ SFR/M$_*$ should be independent of the bulge-to-total ratio.

In Figure \ref{fig:specific} we investigate the relationship between the specific black hole accretion rate (sBHAR) and specific star formation rate (sSFR) for a wide variety of galaxy and AGN type. Black hole masses in this work are based on K-band luminosities of the host galaxy, following \cite{graham07}, while those from literature sources are primarily based on dynamical methods. We find a strong correlation over 6 decades in both sBHAR and sSFR. Unsurprisingly, when the host galaxy is growing most rapidly, the black hole is also growing most rapidly. In general, we find that the central supermassive black hole is growing $\sim$10$\times$ faster than the host galaxy, but we emphasize that these samples are certainly biased towards high BHAR due to being based on detected AGN only -- when non-detections are incorporated, the black holes may actually be growing slower than their host galaxies \citep[e.g.,][]{trump15}. %
%
%
This figure demonstrates that, while we find BHAR/SFR ratios in radio-loud giant elliptical galaxies that are roughly an order of magnitude higher than in typical AGN, this does not necessarily imply dramatically different growth rates. Indeed, Figure \ref{fig:specific} suggests that ``active'' galaxies in general have black holes that are growing at a fairly constant rate, with respect to their host galaxy. The giant elliptical galaxies studied here are growing at a variety of sSFR and sBHAR rates, with a distribution that is lognormal \citep{mcdonald18a}, corresponding to pink noise \citep[$f^{-1}$;][]{gaspari17}. That is, 10\% of the time, the sSFR and sBHAR are elevated by an order of magnitude, while 1\% of the time, they are elevated by two orders of magnitude (e.g., Phoenix, H1821+643, IRAS09104+4109). This figure demonstrates that this chaotic cooling history will not lead to deviations in the M$_{BH}$--M$_{spheroid}$ relations, given that the sSFR/sBHAR ratio remains constant over 4 orders of magnitude in sSFR.

\begin{figure}[t]
\centering
\includegraphics[width=0.47\textwidth]{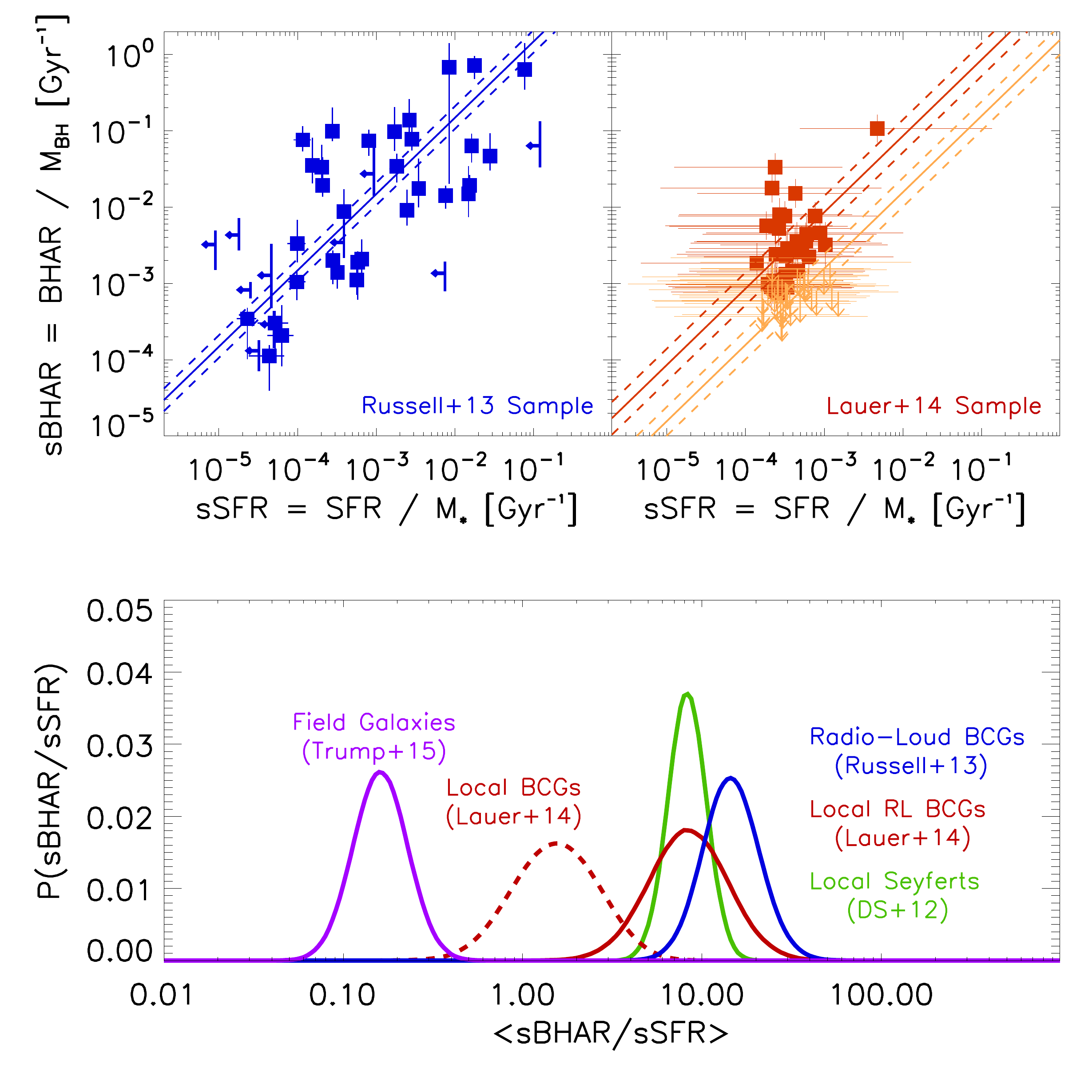}
\caption{This figure shows the constraints on the mean sBHAR/sSFR ratio for a variety of samples. In the upper left panel, we show the main sample from this paper. The best-fit sBHAR/sSFR ratio and its uncertainty (assuming slope unity) is shown. In the upper right we show the unbiased sample of \cite{lauer14}, as discussed in the text. Orange points are non-detections in the radio, while red points are detections. The two lines show the best-fit sBHAR/sSFR ratios including or excluding non-detections. In the lower panel we show the constraints on the sBHAR/sSFR ratio from different samples.
The solid red line shows the constraints on the mean sBHAR/sSFR ratio for the \emph{detected} AGN in the \cite{lauer14} sample, while the dotted red curve shows the constraints when non-detections are included. This plot demonstrates that selecting on detected AGN leads to, roughly, an order-of-magnitude bias in the measured sBHAR/sSFR ratio (consistent with Figures \ref{fig:literature}--\ref{fig:literature_agn}), and that the true underlying sBHAR/sSFR ratio for BCGs is consistent with unity, meaning that black holes and their host galaxies are growing at the same rate, consistent with observations of a universal M$_{\textrm{BH}}$--M$_{\textrm{spheroid}}$ relation.
}
\label{fig:lauer}
\end{figure}

The fact that we find sBHAR/sSFR $\sim$ 10 is certainly a selection effect, since we are considering only active galaxies in Figure \ref{fig:specific}.
To investigate the true underlying ratio of the black hole growth rate to the galaxy growth rate, we consider the complete sample of local BCGs from \cite{lauer14}. We cross-reference the sample of 433 BCGs at $z<0.08$ with the VLA FIRST survey \citep{becker95} and the Sloan Digital Sky Survey \citep[SDSS DR8;][]{aihara11} Value Added Catalog of \cite{brinchmann04}, to acquire 1.4 GHz radio luminosities (or upper limits), star formation rates, and stellar masses for 68 BCGs that have observations (though not necessarily detections) in all three surveys. Assuming that the spectroscopic follow-up of SDSS is not a function of BCG star formation rate or radio power, this should still represent an unbiased sample. We convert radio power to cavity power following \cite{cavagnolo10}, and assume a M$_{BH}$/M$_*$ ratio consistent with the median value for the \cite{russell13} BCGs, to allow for a fair comparison ($\left<\textrm{M}_{BH}/\textrm{M}_*\right> = 0.002$).
In Figure \ref{fig:lauer}, we compare the mean ratio of the specific BHAR to specific SFR (sBHAR/sSFR) for several samples. This is equivalent to the normalization of the relationship shown in Figure \ref{fig:specific}, assuming slope unity. To incorporate non-detections, we assume an underlying log-normal distribution of sBHAR/sSFR, and follow the methodology of \cite{kelly07}, fitting only for the normalization. We find that, when considering only detected AGN, the mean sBHAR/sSFR ratio is $\sim$10, consistent with Figure \ref{fig:specific}. Indeed, whether we consider the radio-loud BCGs in the sample of \cite{lauer14}, the mechanically-powerful AGN in the sample of \cite{russell13}, or the sample of local Seyfert galaxies from \cite{diamond-stanic12}, we measure a consistent value of $\left<\textrm{sBHAR/sSFR}\right>$ to within the uncertainties. However, when non-detections are incorporated in the \cite{lauer14} sample, we find $\left<\textrm{sBHAR/sSFR}\right> \sim 1$, which is much closer to the unbiased estimate for field galaxies from \cite{trump15}.  This exercise confirms that, while the typical AGN-selected sample will have $\left<\textrm{sBHAR/sSFR}\right> \sim 10$ (Figure \ref{fig:specific}), the underlying population of galaxies has $\left<\textrm{sBHAR/sSFR}\right> \sim 1$, which is consistent with the observed scaling relations between black hole and host galaxy mass \citep[e.g.,][]{kormendy13}. 

The factor of $\sim$10 correction when non-detections are included would bring the measured $\log_{10}$(BHAR/SFR) of $-1.6$ from this work in line with the measurement of $-2.5$ from \cite{yang19b} for spheroidal galaxies. Moreover, it would support the results of \cite{yang18} who predicted that galaxies with M$_*$ $\sim$ 10$^{12}$ M$_{\odot}$ should have BHAR/SFR $\sim$ 10$^{-3}$, based on an extrapolation of the relation between BHAR/SFR and M$_*$. This consistency (see also Figure \ref{fig:mstar}) strongly supports a picture where the most massive galaxies in the universe are, on average, growing their black holes at a faster rate than their lower-mass, disk-dominated peers.

In summary, while we find evidence that the BHAR/SFR ratio is elevated in giant elliptical galaxies compared to the general population, we do not find that this is in tension with the observations of a single M$_{BH}$--M$_{spheroid}$ relation. In order to have a universal M$_{BH}$--M$_{spheroid}$ relation, purely-spheroidal systems must be growing their black holes at a higher rate per unit total stellar mass than disk-dominated systems. By normalizing out the total stellar and black hole masses, we remove this morphological dependence, showing that all active galaxies sit on a single sBHAR--sSFR relation, with \emph{active} black holes growing, on average, $\sim$10$\times$ faster than their host galaxies. This factor of $\sim$10 is due to selecting on active galaxies -- if we consider the whole galaxy population, we find no evidence that black holes and their host galaxies are growing at diverging rates.

\section{Summary}

In this work, we have considered the ratio of the black hole accretion rate (BHAR) to the star formation rate (SFR) for a sample of giant elliptical galaxies.  These galaxies, which tend to live at the centers of massive groups and clusters of galaxies, are predominantly radio loud and are responsible for regulating cooling of the intracluster medium. For the first time, we consider the relationship between BHAR and SFR where the BHAR is derived based on the jet power of the central AGN. Utilizing these data, along with a rich selection of data from the literature, we arrive at the following conclusions:

\begin{itemize}

\item We find a strong correlation between BHAR and SFR for radio-loud giant ellipitical galaxies, where the former is constrained based on the jet power of the central AGN. The mean BHAR/SFR ratio for these galaxies is $\log_{10}(\textrm{BHAR/SFR}) = -1.45 \pm 0.2$. This measurement is independent of the star formation indicator (considering both H$\alpha$ and M$_{H_2}$ as proxies for ongoing star formation) and independent of the methodology of constraining the BHAR (P$_{cav}$, L$_{\textrm{X}}$, $\dot{M}_{\textrm{Bondi}}$). This high BHAR/SFR ratio is consistent with what is needed for mechanical AGN power to offset cooling of the intracluster medium and suppress star formation.

\item The mean BHAR/SFR ratio measured here for giant elliptical galaxies appears to be roughly an order of magnitude higher than what is typically quoted in the literature. Literature estimates span $-4< \log_{10}(\textrm{BHAR/SFR}) < -2$. Several studies have shown that this ratio is dependent on the host mass, which we confirm here for the most massive galaxies.

\item Considering a variety of samples from the literature, we find that the mean BHAR/SFR ratio correlates most strongly with host galaxy type (i.e., morphology), with spheroidal galaxies having consistently higher BHAR/SFR ratios than disk or star-forming galaxies. There is some evidence for higher BHAR/SFR in radio galaxies, but we suspect that this is a secondary effect and that galaxy morphology is the primary driver of the scatter. We find no evidence that the measured BHAR/SFR ratio correlates with distance from the main sequence of star formation. 

\item We propose that the link to host galaxy morphology is related to angular momentum, with spheroidal galaxies having more efficient accretion onto their central black holes due to cool clouds having predominantly radial trajectories as they condense out of the hot halo \citep[i.e., via Chaotic Cold Accretion;][]{gaspari13b,gaspari15b}, while disk galaxies have the bulk of their cool gas in a disk, where star formation is efficient and accretion onto the central black hole is relatively suppressed. The longer timescales for gas to travel from large radii to small in disk galaxies should yield overall lower BHAR/SFR ratios, as more gas is consumed along the way (compared to cooling on nearly-radial trajectories).

\item We find a strong correlation between the \emph{specific} BHAR and SFR (sBHAR and sSFR) over six orders of magnitude in both parameters. This correlation is consistent with opposite dependencies on the bulge-to-total ratio between the M$_{\textrm{BH}}$--M$_{\textrm{*}}$ relation and the BHAR--SFR relation. We find that most AGN, detected via a variety of different methods, have sBHAR/sSFR $\sim$10, meaning that their black holes are growing more rapidly than the host galaxy. We demonstrate that this is driven by selecting on active galaxies -- for a complete sample of local BCGs, including those that are not detected in the radio, we find $\textrm{sBHAR/sSFR} \sim 1$, consistent with a universal M$_{\textrm{BH}}$--M$_{\textrm{spheroid}}$ relation.

\end{itemize}

While this study implies a link between the BHAR/SFR ratio and the host galaxy morphology, we have not definitively demonstrated this. It could very well be that radio power scales differently with accretion rate than does accretion luminosity, as we discuss in \S5.3. The next step would be to assemble a complete sample spanning a broad range in galaxy type, galaxy mass, AGN type, and galaxy environment, and to determine which is the primary driver for the scatter in BHAR/SFR. With the era of large, multiwavelength surveys upon us, such analyses are entirely realistic and can be pursued right away. Further, simulations can more readily address the role of angular momentum in governing the BHAR/SFR ratio, and whether this is the primary driver of the scatter that we observe.

%
%
\section*{Acknowledgements} 

We thank the anonymous referee for their careful reading of this manuscript and their thoughtful, constructive feedback.
This research has made use of the NASA/IPAC Extragalactic Database (NED), which is funded by the National Aeronautics and Space Administration and operated by the California Institute of Technology.
This work is based in part on observations with the NASA/ESA/CSA Hubble Space Telescope obtained at the Space Telescope Science Institute, which is operated by the Association of Universities for
Research in Astronomy, Incorporated, under NASA contract NAS5-26555. Support for Program number HST-GO-15661.001-A was provided through a grant from the STScI under NASA contract NAS5-26555. MM received additional support from the National Science Foundation through CAREER award number AST-1751096.
R.C.H. acknowledges support from NASA through grant number 80NSSC19K0580, and the National Science Foundation through CAREER award number 1554584.
MG acknowledges partial support by the Lyman Spitzer Jr.\ Fellowship (Princeton University) and by NASA Chandra GO8-19104X/GO9-20114X and HST GO-15890.020-A grants.
%


\end{document}